\documentclass[12pt]{article}
\pdfoutput=1
\usepackage[utf8]{inputenc}
\usepackage{amsfonts,amsmath}
\usepackage{jheppub}

\usepackage{color}
\usepackage{tabularx}

\usepackage{bbm}

\usepackage{tikz}
\usetikzlibrary{positioning,arrows}
\usetikzlibrary{decorations.pathmorphing,calc}
\usetikzlibrary{decorations.markings}
\newif\ifmirrorsemicircle
\usepackage{float}
\usepackage[justification=centering]{caption}

\usepackage{ifthen}
\usepackage{enumerate}
\usepackage{amsmath}
\usepackage{amssymb}
\usepackage[mathscr]{eucal}
\usepackage[errorshow]{tracefnt}
\usepackage{amsmath}
\usepackage{slashed}
\usepackage{amssymb}
\usepackage{array}
\usepackage{amsthm}
\usepackage{eucal}
\usepackage{epsf}
\usepackage{epsfig}
\usepackage{psfrag}
\usepackage{multirow}
\usepackage{wasysym} 
\usepackage{tikz-cd} 
\usepgfmodule{shapes}
\usetikzlibrary{backgrounds, arrows,calc,shapes,decorations.pathreplacing, decorations.markings, automata,positioning}

\newif\ifmirrorsemicircle

 \definecolor{mygreen}{RGB}{50,205,50}
 
 \makeatletter
\newcommand*\bigcdot{\mathpalette\bigcdot@{.5}}
\newcommand*\bigcdot@[2]{\mathbin{\vcenter{\hbox{\scalebox{#2}{$\m@th#1\bullet$}}}}}
\makeatother

\newcommand*\DAlambert{\mathop{}\!\mathbin\Box}

\tikzset{
    partial ellipse/.style args={#1:#2:#3}{
        insert path={+ (#1:#3) arc (#1:#2:#3)}}}

\tikzset{
    wave amplitude/.initial=0.07cm,
    wave count/.initial=7,
    mirror semicircle/.is if=mirrorsemicircle,
    mirror semicircle=false,
    wavy semicircle/.style={
        to path={
            let \p1 = (\tikztostart),
            \p2 = (\tikztotarget),
            \n1 = {veclen(\x2-\x1,\y2-\y1)},
            \n2 = {atan2(\x2-\x1,\y2-\y1))} in
            plot [
                smooth,
                samples=(\pgfkeysvalueof{/tikz/wave count}+0.5)*8+1, 
                domain=0:1,
                shift={($(\p1)!0.5!(\p2)$)}
            ] ({ 
                (\x*180-\n2 + 180 + \ifmirrorsemicircle 1 \else -1 \fi * 90%
            }:{ 
                (%
                    \n1/2+\pgfkeysvalueof{/tikz/wave amplitude} * %
                    sin(
                        \x * 360 * (\pgfkeysvalueof{/tikz/wave count} + 0.5%
                    )%
                )%
            })
        } (\tikztotarget) }}

\pgfdeclaredecoration{complete sines}{initial}
{
    \state{initial}[
        width=+0pt,
        next state=sine,
        persistent precomputation={\pgfmathsetmacro\matchinglength{
            \pgfdecoratedinputsegmentlength / int(\pgfdecoratedinputsegmentlength/\pgfdecorationsegmentlength)}
            \setlength{\pgfdecorationsegmentlength}{\matchinglength pt}
        }] {}
    \state{sine}[width=\pgfdecorationsegmentlength]{
        \pgfpathsine{\pgfpoint{0.25\pgfdecorationsegmentlength}{0.5\pgfdecorationsegmentamplitude}}
        \pgfpathcosine{\pgfpoint{0.25\pgfdecorationsegmentlength}{-0.5\pgfdecorationsegmentamplitude}}
        \pgfpathsine{\pgfpoint{0.25\pgfdecorationsegmentlength}{-0.5\pgfdecorationsegmentamplitude}}
        \pgfpathcosine{\pgfpoint{0.25\pgfdecorationsegmentlength}{0.5\pgfdecorationsegmentamplitude}}
}
    \state{final}{}
}

\tikzset{ 
    fermion/.style={thick, draw=mygreen, postaction={decorate}, decoration={markings, mark=at position .5 with {\arrow[mygreen]{triangle 45}}}} ,
       scalar/.style={thick, draw=blue, postaction={decorate}, decoration={markings, mark=at position .43 with {\arrow[blue]{triangle 45}}}} ,
          scalarr/.style={thick, draw=blue, postaction={decorate}, decoration={markings, mark=at position .56 with {\arrowreversed[blue]{triangle 45}}}},
       photon/.style={decorate, draw=black,thick,
        decoration={complete sines,amplitude=5pt, segment length=10pt}},
         photon1/.style={decorate, draw=black,thick,
        decoration={complete sines,amplitude=4pt, segment length=7pt}},
        photon2/.style={decorate, draw=black,thick,
        decoration={complete sines,amplitude=2.5pt, segment length=7pt}},
        photon3/.style={decorate, draw=black,thick,
        decoration={complete sines,amplitude=2.5pt, segment length=6pt}},
        photon0/.style={decorate, draw=black,thick,
        decoration={complete sines,amplitude=5pt, segment length=7pt}},
         photon01/.style={decorate, draw=black,thick,
        decoration={complete sines,amplitude=3pt, segment length=5pt}}
      }

\tikzset{/pgf/decoration/.cd,
    number of sines/.initial=10,
    angle step/.initial=20,
}
\newdimen\tmpdimen
\pgfdeclaredecoration{complete siness}{initial}
{
    \state{initial}[
        width=+0pt,
        next state=move,
        persistent precomputation={
            \pgfmathparse{\pgfkeysvalueof{/pgf/decoration/angle step}}%
            \let\anglestep=\pgfmathresult%
            \let\currentangle=\pgfmathresult%
            \pgfmathsetlengthmacro{\pointsperanglestep}%
                {(\pgfdecoratedremainingdistance/\pgfkeysvalueof{/pgf/decoration/number of sines})/360*\anglestep}%
        }] {}
    \state{move}[width=+\pointsperanglestep, next state=draw]{
        \pgfpathmoveto{\pgfpointorigin}
    }
    \state{draw}[width=+\pointsperanglestep, switch if less than=1.25*\pointsperanglestep to final, 
        persistent postcomputation={
        \pgfmathparse{mod(\currentangle+\anglestep, 360)}%
        \let\currentangle=\pgfmathresult%
    }]{%
        \pgfmathsin{+\currentangle}%
        \tmpdimen=\pgfdecorationsegmentamplitude%
        \tmpdimen=\pgfmathresult\tmpdimen%
        \divide\tmpdimen by2\relax%
        \pgfpathlineto{\pgfqpoint{0pt}{\tmpdimen}}%
    }
    \state{final}{
        \ifdim\pgfdecoratedremainingdistance>0pt\relax
            \pgfpathlineto{\pgfpointdecoratedpathlast}
        \fi
   }
}

\newcommand{\bC}{\mathbb{C}}
\newcommand{\bP}{\mathbb{P}}

\newcommand{\bpic}{\begin{tikzpicture}}
\newcommand{\epic}{\end{tikzpicture}}
\newcommand{\ba}{\begin{array}}
\newcommand{\ea}{\end{array}}

\title{Higher Derivative Gauge theory in $d=6$ and the $\mathbb{C}\mathbb{P}^{(N_f-1)}$ NLSM}

\preprint{}

\author[]{Hrachya Khachatryan}

\affiliation[]{International School of Advanced Studies (SISSA), Via Bonomea 265, 34136 Trieste, Italy}

\emailAdd{hkhachat@sissa.it}

\abstract{
 We consider the $\mathbb{C}\mathbb{P}^{(N_f-1)}$  Non-Linear-Sigma-Model in the dimension $4<d<6$. The critical behaviour of this model in the large $N_f$ limit is reviewed. We propose a Higher Derivative Gauge (HDG) theory as an ultraviolet completion of the $\mathbb{C}\mathbb{P}^{(N_f-1)}$  NLSM. Tuning mass operators to zero, the HDG in the IR limit reaches to the critical $\mathbb{C}\mathbb{P}^{(N_f-1)}$. With partial tunings the HDG reaches either to the critical $U(N_f)$-Yukawa model or to the critical pure scalar QED (no Yukawa interactions). 

We renormalize the HDG in its critical dimension $d=6$. We study the fixed points of the HDG in $d=6-2\epsilon$ and we calculate the scaling dimensions of various observables finding a full agreement with the order $O(1/N_f)$ predictions  of the corresponding critical models. 
}

\begin{document}

\maketitle
\newpage

\section{Introduction and Summary}
In the paper \cite{Fei:2014yja}, Fei, Giombi and Klebanov studied the $O(N)$ vector model in the dimension $4<d<6$. We remind that the d-dimensional $O(N)$ vector model with $N$ real scalars $\phi_i$, is defined with the following action
  \begin{align} \label{I1}
 S_{O(N)}=\int d^dx \Big[ \frac{1}{2} (\partial_\mu \phi_i)^2+\lambda \big( \sum\limits_{i=1}^N \phi_i^2\big)^2 \Big] \ .
 \end{align}
 Notice that the mass term $\tau \sum\limits_{i=1}^N \phi^2_i$ is tuned to zero or equivalently the temperature T is tuned to its critical value $T_c$ ($\tau=\frac{T-T_c}{T_c} \rightarrow 0$). Using the Hubbard-Stratonovich (HS) transformation  \cite{Stratonovich, Hubbard} one is able to trade the quartic interaction  with cubic and quadratic terms.  Then the vector model (\ref{I1}) is described by the equivalent theory   
  \begin{align} \label{I2}
  S=\int d^d x \Big[ (\partial \phi_i)^2+\sigma\phi_i^2 -\frac{\sigma^2}{4\lambda} \Big] \ ,
  \end{align}
  where the summation over the flavor index $i=1,2,...,N$ is assumed. Indeed integrating out the scalar (HS) field $\sigma$ one will obtain the  $O(N)$ vector model (\ref{I1}). 
  
  Let us review the large $N$ limit of the $O(N)$ vector model \cite{Domb}. The graphs that contribute to the 2-point correlation function of the HS field are the bubble graphs in (\ref{I3}), all other graphs are $1/N$ suppressed. 
  
   \begin{equation} \label{I3}
  \begin{tikzpicture} [scale=0.59]  
  \node at (0,0) {$\langle \sigma(p) \sigma(-p) \rangle$};
  \node at (2.1,-0.05) {$ \ \ =$};
  \draw[dashed,thick] (3,0) to (5,0);
  \node at (5.5,0) {$+$};
  \draw[dashed,thick] (6,0) to (7.1,0);
  \draw[thick] (8.1,0) circle(1cm);
   \draw[dashed,thick] (9.1,0) to (10.2,0);
  \node at (10.7,0) {$+$};
   \draw[dashed,thick] (11.3,0) to (12.4,0);
  \draw[thick] (13.4,0) circle(1cm);
   \draw[dashed, thick] (14.4,0) to (15.5,0);
   \draw[thick] (16.5,0) circle (1cm);
   \draw[dashed,thick] (17.5,0) to (18.6,0);
   \node at (19.5,0) {$ \ \ \  \ \ \ + \  \ .... \ \ \ ,$};     
  \end{tikzpicture}
  \end{equation}
  
  The dashed line in the graphs (\ref{I3}) represents  the  ``tree level" propagator of the HS field. For a single bubble graph we have 
 \begin{align} \label{I4}
 2N  \ \int \frac{d^d q}{(2\pi)^d} \frac{1}{q^2 (p-q)^2} = N \frac{2 \Gamma(\frac{d}{2}-1)^2 \Gamma(2-\frac{d}{2})}{(4\pi)^{d/2} \Gamma(d-2)} p^{d-4}=N A(d) p^{d-4} \ ,
 \end{align}
 where the factor $N$ is due to the $N$ scalars $\phi_i$ circulating inside the closed loop (\ref{I3}). The fraction in (\ref{I4}) is denoted by $A(d)$. To calculate the integral we used (\ref{ma2}). Summing geometric series of the bubble graphs in (\ref{I3}) gives
  \begin{align} \nonumber
  \langle \sigma(p) \sigma(-p)\rangle &= (-4\lambda)+ (-4\lambda) N A(d) p^{d-4} (-4\lambda)+ (-4\lambda) \big(N A(d) p^{d-4} (-4\lambda)\big)^2+... \\
  &=(-4\lambda) \frac{1}{1+4\lambda N A(d) p^{d-4}} \label{I5} \ .
  \end{align}
  It follows from (\ref{I5}), that in the dimension $2<d<4$, the HS field $\sigma$ in the IR limit $p\rightarrow 0 $ has the following scaling  
  \begin{align} \label{I6}
   \langle \sigma(p) \sigma(-p)\rangle|_{p\rightarrow 0} =(-4\lambda) \frac{1}{1+4\lambda N A(d) p^{d-4}} \bigg|_{p\rightarrow 0}=-\frac{p^{4-d}}{N A(d)} \ .
  \end{align}
  We conclude that in the  large $N$ limit, at the IR critical point $\Delta[\sigma]=2$. Therefore the $\sigma^2$ operator has a scaling dimension equal to 4 and it is an irrelevant operator. For the $O(N)$ vector model, the NLO $O(1/N)$ and the NNLO $O(1/N^2)$ corrections to the scaling dimension of various observables have been studied in \cite{Vasiliev:1981yc, Vasiliev:1981dg}. The scaling dimension of the scalar field $\phi_i$  was calculated at the order $O(1/N^3)$ in \cite{Vasiliev:1982dc}, using the conformal bootstrap equations. 
  
    When $d>4$ the $\phi^4$ operator\footnote{Shorthand notation $\phi^4=\big(\sum\limits_{i=1}^{N} \phi_i^2\big)^2$.}  is an irrelevant deformation, and in \cite{Fei:2014yja} (see also \cite{Parisi:1975im}) the existence of an UV interacting fixed point  was conjectured\footnote{See however \cite{Percacci:2014tfa}, where the authors using the functional RG seem to rule out existence of such a UV interacting fixed point with bounded critical potential. See also \cite{Mati:2014xma,Mati:2016wjn}. In particular in the last reference, bounded but non-analytic critical potential has been found.}. Indeed it follows from (\ref{I6}), that in the dimension $4<d<6$, the 2-point function of the HS field scales as $\sim p^{4-d}$ when $p\rightarrow \infty$ (UV limit).  Notice that in contrast to the case $d<4$, in $d>4$ the operator $\sigma^2$ in the large $N$ limit  is a relevant operator at the critical point (since it has a scaling dimension  equal to $4>d$). 
  The theory (\ref{I2}) was UV completed in $4<d<6$ \cite{Fei:2014yja}: including in the action a kinetic term $(\partial_\mu\sigma)^2$ and a cubic term $\sigma^3$. 
   Because of the presence of a ``Yukawa" type interaction $\sigma\phi^2$, we will refer to this model as $O(N)$-Yukawa. It is very crucial to observe that these ultraviolet completion in the dimension $4<d<6$ has a relevant operator $\sigma^2$, which must be  tuned to zero (the mass term $\phi^2$ must be tuned to zero as well) in order to reach the IR critical point. The IR critical $O(N)$-Yukawa model was identified with the UV interacting fixed point of the $O(N)$ vector model in $4<d<6$. 
       
 The $O(N)$-Yukawa model was examined  near its critical dimension $d=6$ \cite{Fei:2014yja} (the critical dimension of a given theory is defined as the dimension where the interactions in the action become marginal). In its critical dimension the theory was renormalized at one loop (later 3-loop \cite{Fei:2014xta} and four loop \cite{Gracey:2015tta} analysis have been carried out). It was proved that the IR stable interacting fixed point at $d=6-2\epsilon$ coincides with the critical $O(N)$ vector model. 
   
 Motivated with this discussion, in this paper  we study the $\bC\bP^{(N_f-1)}$ NLSM in $4<d<6$ with $N_f$ complex fields $\Phi_i$.  This model will be engineered with the help of two master (HS) fields: the vector $A_\mu$ and the scalar $\sigma$.  The operators $(\sigma^2, F^2_{\alpha\beta})$  are relevant at the critical point in the large $N_f$ limit, since both have scaling dimension $4>d$ (see section (\ref{largeNf}) for mode details). 
  
  The $\bC\bP^{(N_f-1)}$ model engineered with the help of two master fields, will be UV completed including in the action the ``kinetic terms": $ (\partial_\mu \sigma)^2, (\partial_\mu F_{\alpha\beta})^2 $ and the interaction terms: $\sigma^3, \sigma F_{\alpha\beta}^2$. Notice that the kinetic term of the gauge field contains 4-derivatives, instead the term $F_{\alpha\beta}^2$ plays a role of a gauge invariant mass term for the gauge field. For this reason we will refer to the UV completion as a Higher Derivative Gauge (HDG) theory. In this theory the mass terms $(\sigma^2, F_{\alpha\beta}^2)$ are relevant deformations, and we need to tune both of them to zero in order to reach the critical $\bC\bP^{(N_f-1)}$. If we choose to not tune the $\sigma^2$ term, then we end up on another interesting critical point: the critical pure scalar QED in $4<d<6$ (no Yukawa interaction of type $\sigma\Phi^2$). Notice that in the dimension $4<d<6$ the operator $\Phi^4$ is irrelevant, and therefore to reach the IR critical scalar QED, there will be no need to tune that operator to zero (this was not the case in $2<d<4$, where for instance to reach the tricritical point we have to tune to zero the quartic operator). Instead if we do not tune the term $F_{\alpha\beta}^2$, then the RG will end on the critical $O(N)$-Yukawa, which has already been discussed in \cite{Fei:2014yja}. 
  
  Most importantly we renormalize the HDG in its critical dimension $d=6$. In the dimension $d=6-2\epsilon$ (taking $N_f$ large) we find two IR interacting fixed points (besides the ungauged fixed point which corresponds to the critical $O(N)$-Yukawa). We prove that these fixed points coincide with the critical $\bC\bP^{(N_f-1)}$ and the critical pure scalar-QED. 
  
We also mention that in \cite{Percacci:2014tfa, Mati:2014xma, Mati:2016wjn, 
Giombi:2015haa, Gracey:2015xmw, Herbut:2015zqa, Eichhorn:2016hdi, Brust:2016gjy, Roscher:2018ucp, Gracey:2018khg} the vector model, tensor models, fermionic QED and fermionic QCD have been studied in the dimension $4<d<6 $.  Also see
 \cite{Kazakov:2002jd, Ivanov:2005qf, Bossard:2015dva, Casarin:2019aqw}, where the supersymmetric higher derivative gauge theories have been considered. 
  
  The paper is organized as follows. First we review the model $\bC\bP^{(N_f-1)}$ and its properties at the large $N_f$ limit in $4<d<6$. In particular we provide scaling dimensions of various operators at the order $O(1/N_f)$ in d-dimension. 
  We also discuss the critical scalar-QED in the large $N_f$ limit.  The large $N_f$ limit of this model has not been studied yet in the literature, we provide scaling dimensions of some operators without giving the details of the computations. Second, we renormalize the UV action at $d=6$ by constructing the one-loop beta functions, one-loop anomalous dimensions of the fields and of the mass operators (mass renormalization). The beta functions are solved in the large $N_f$ limit and the fixed points are classified. At all fixed points the scaling dimensions of the fields and of the mass operators are explicitly provided. Finally, these results are checked versus the large $N_f$ predictions of the critical models.

  \section{Large $N_f$ expansion of the critical $\bC\bP^{(N_f-1)}$ NLSM } \label{largeNf}
  
  The $\bC\bP^{(N_f-1)}$ Non-Linear-Sigma-Model is described by $N_f$ complex scalar fields subject to the condition $\sum\limits_{i=1}^{N_f} |\Phi_i|^2=N_f$, with the following action
  \begin{align} \label{vas1}
  S_{\bC\bP{(N_f-1)}}=\int d^dx \Big[ \sum\limits_{i=1}^{N_f} |\partial_\mu \Phi_i|^2+\frac{1}{4N_f}  \big(\sum\limits_{i=1}^{N_f}( \Phi_i^* \partial_\mu \Phi_i-\partial_\mu \Phi_i^* \cdot \Phi_i )\big)^2 \Big] \ .
  \end{align}
  The action is easily proved to be gauge invariant under the local $U(1)$ transformations $\Phi_i(x)\rightarrow e^{i\alpha(x)} \Phi_i(x)$. Due to the constraint $\sum\limits_{i=1}^{N_f} |\Phi_i|^2=N_f$, the vector $\Phi_i$ lies on a sphere $S^{2N_f-1}$. Additionally the gauge invariance implies that the field configurations related by the gauge transformations are physically equivalent, and inside the path integral one shouldn't integrate over these equivalent configurations. Geometrically this means that the target space becomes $\bC\bP^{(N_f-1)} \sim S^{2N_f-1}/U(1)$. 
  
  As it is usually the case, for building the $1/N_f$ expansion it is comfortable to introduce master (HS) fields: a scalar field $\sigma$ as a Lagrange multiplier for the constraint, and a vector field $A_\mu$ to engineer the complicated quartic interaction with derivatives of (\ref{vas1}) as a sum of quadratic and cubic terms (Hubbard-Stratonovich transformation). This allows to rewrite the action (\ref{vas1}) as follows
  \begin{align} \label{vas2}
  S=\int d^d x \big[ \sum\limits_{i=1}^{N_f} |\partial_\mu \Phi_i|^2+iA_\mu \sum\limits_{i=1}^{N_f}( \Phi_i^* \partial_\mu \Phi_i-\partial_\mu \Phi_i^* \cdot \Phi_i ) +N_f A_\mu^2+\sigma \big(\sum\limits_{i=1}^{N_f} |\Phi_i|^2-N_f\big) \big] \ .
  \end{align}
  After shifting $\sigma\rightarrow \sigma+A_\mu^2$ the action (\ref{vas2}) takes the following simple form
  \begin{align} \label{vas3}
   S=\int d^d x \big[ \sum\limits_{i=1}^{N_f} |D_\mu \Phi_i|^2+\sigma \big(\sum\limits_{i=1}^{N_f} |\Phi_i|^2-N_f\big) \big] \ .
  \end{align}
  The $U(1)$ gauge invariance of (\ref{vas3}) is obvious, with the vector field $A_\mu$ playing the role of the gauge field. The gauge fixing term is required to fix the redundancies, following \cite{Vas:1983} the standard $\mathbbm{R}_\xi$ gauge is employed ($\xi=0$ is the Landau gauge). From now on we are interested in the large $N_f$ limit of the (\ref{vas3}).
  
  In \cite{Vas:1983} Vasil'ev and Nalimov studied (\ref{vas3}) in the dimension $2<d<4$ \footnote{The non-abelian generalization of the $\bC\bP^{(N_f-1)}$ model was discussed in \cite{Vasiliev:1984bf}.}, see also \cite{Hikami:1979ih}. They calculated, at the critical point, the leading order scaling dimensions of the master fields in the large $N_f$ limit: $\Delta[\sigma]=2, \ \Delta[A_\mu]=1$. They also observed that the scalar QED in $2<d<4$ is in the same universality class with the $\bC\bP^{(N_f-1)}$ NLSM. The NLO corrections to the scaling dimensions of various observables were also calculated in \cite{Vas:1983}. We will summarize their results at the end of this section. Before proceeding,  we briefly remind why in the scalar QED ($2<d<4$)  in the large $N_f$ limit $\Delta[A_\mu]=1$. The scalar QED action (after applying the HS transformation on $\Phi^4$ interaction) is defined by  (\ref{vas3}), adding to it a kinetic term for the photon $F_{\mu\nu}^2/4e^2$. In the large $N_f$ limit, only the following bubble graphs contribute to the 2-point function of the photon.
    \begin{equation} \label{effective photon}
  \begin{tikzpicture} [scale=0.57]  
  \node at (-0.9,0) {$\langle A_\mu(p) A_\nu(-p) \rangle$};
  \begin{scope}[shift={(-0.3,0)}]
  \node at (2.1,-0.05) {$ \ \ =$};
  \draw[photon1] (3,0) to (5,0);
  \node at (5.5,0) {$+$};
  \draw[photon1] (6,0) to (7.1,0);
  \draw[thick] (8.1,0) circle(1cm);
   \draw[photon1] (9.1,0) to (10.2,0);
  \node at (10.7,0) {$+$};
   \draw[photon1] (11.3,0) to (12.4,0);
  \draw[thick] (13.4,0) circle(1cm);
   \draw[photon1] (14.4,0) to (15.5,0);
   \draw[thick] (16.5,0) circle (1cm);
   \draw[photon1] (17.5,0) to (18.6,0);
   \node at (19.5,0) {$ \ \  \ \ \ + \  \ .... \ \ ,$};     
   \end{scope}
  \end{tikzpicture}
  \end{equation}
   The wavy line in the graphs (\ref{effective photon}) represents  the  tree level photon propagator in the Landau gauge $D_{\alpha\beta}(p)=\frac{e^2}{p^2} \big(\delta_{\alpha\beta} -\frac{p_\alpha p_\beta}{p^2} \big)$. For a single bubble graph we have 
 \begin{align} \nonumber 
 &\Pi_{\alpha\beta}(p)=N_f  \ \int  \frac{d^d q}{(2\pi)^d} \frac{(2q+p)_\alpha (2q+p)_\beta}{q^2 (p+q)^2} \\
 &= N_f \frac{2^{2-d} \sqrt{\pi} \Gamma(1-d/2)\Gamma(d/2)}{(4\pi)^{d/2} \Gamma(d/2+1/2)} \big(\delta_{\alpha\beta}-\frac{p_\alpha p_\beta}{p^2}\big) p^{d-2} 
 =N_f B(d) \big(\delta_{\alpha\beta}-\frac{p_\alpha p_\beta}{p^2}\big) p^{d-2} \ , \label{vas20}
 \end{align}
 where the factor $N_f$ is due to  $N_f$ complex scalar flavors circulating inside the closed loop (\ref{effective photon}). The fraction in (\ref{vas20}) is denoted by $B(d)$. Summing geometric series of the bubble graphs in (\ref{effective photon}) gives
  \begin{align} \label{vas23}
  \langle A_\mu(p) A_\nu(-p)\rangle = D_{\mu\rho} (1-\Pi D)^{-1}_{\rho\nu}= \frac{e^2}{p^2} \big(\delta_{\mu\nu} -\frac{p_\mu p_\nu}{p^2} \big) \frac{1}{1-N_f B(d) e^2 p^{d-4}} \ .
  \end{align}
  Therefore  we conclude that in $2<d<4$, the scaling dimension of the photon  in the IR limit is $\Delta[A_\mu]=1$ \footnote{See also \cite{Benvenuti:2019ujm}, where a similar proof was carried out for scalar QED in d=3.},
  \begin{align} \label{vas22}
   \langle A_\mu(p) A_\nu(-p)\rangle|_{p\rightarrow 0} =-\frac{\big(\delta_{\mu\nu}-\frac{p_\mu p_\nu}{p^2}\big) p^{2-d}}{N_f B(d) } \ .
     \end{align}
  
  We are interested to examine (\ref{vas3}) in the dimension $4<d<6$. The analysis made in \cite{Vas:1983} still holds, in particular we can use their results by simply analytically continuing the dimension $d$.  However there is one crucial difference: in $d<4$ the critical $\bC\bP^{(N_f-1)}$ is realized as an IR fixed point of the scalar QED (plut $\Phi^4$ interaction), while in $d>4$ it is an UV interacting fixed point of that theory. Indeed, it follows from (\ref{vas23}) that for $d>4$, one recovers the scaling behaviour (\ref{vas22}) when $p\rightarrow \infty$ (UV limit).
  
   We propose a Higher Derivative Gauge (HDG) theory as a UV completion of the action (\ref{vas3}): including in (\ref{vas3}), the kinetic terms $ (\partial_\mu \sigma)^2, (\partial_\mu F_{\alpha\beta})^2 $ and the interaction terms $\sigma^3, \sigma F_{\alpha\beta}^2$. In the next section we will see that the HDG is asymptotically free. Besides to the above mentioned terms, one can also include in the HDG action  ``mass" terms $(\sigma^2, F_{\alpha\beta}^2)$, which are relevant deformations. Tuning to zero these terms (also the $\Phi^2$ term), the HDG in the IR limit  flows to the critical $\bC\bP^{(N_f-1)}$.  Indeed, below  we will show that in the large $N_f$ limit the IR scaling dimensions are $\Delta[\sigma]=2, \ \Delta[A_\mu]=1$. Therefore the operators $(\partial_\mu \sigma)^2, (\partial_\mu F_{\alpha\beta})^2,\sigma^3, \sigma F_{\alpha\beta}^2$ are irrelevant at the critical point and the HDG in the IR limit is  effectively described by (\ref{vas3}). 
   
    Let us check the  statement $\Delta[A_\mu]=1$ (similarly one can check that tuning $\sigma^2$  to zero, in large $N_f$ limit $\Delta[\sigma]=2$). Since we tuned the ``mass" term $F_{\alpha\beta}^2$ to zero, the tree level propagator for the photon is solely determined by the higher-derivative kinetic term, which gives $D(p)\sim \frac{1}{p^4}$ (see more details in the next section). Repeating the steps of (\ref{vas20}), with that tree level propagator one obtains
  \begin{align}
  \langle A_\mu(p) A_\nu(-p)\rangle= \frac{e^2}{p^4} \big(\delta_{\mu\nu} -\frac{p_\mu p_\nu}{p^2} \big) \frac{1}{1-N_f B(d) e^2 p^{d-6}}  \ .\label{vas21}
  \end{align}
  From (\ref{vas21}) it follows that in the IR limit $p\rightarrow 0$, one recovers the behaviour (\ref{vas22}), which proves $\Delta[A_\mu]=1$.

   In the case, when the $F_{\alpha\beta}^2$ is turned on, in the IR limit we end up on the critical  $U(N_f)$-Yukawa. Instead, when the $\sigma^2$ is turned on, in the IR limit we end up on the critical scalar-QED (one may call it a pure scalar QED, since the Yukawa interactions $\sigma\Phi^2$ are absent). Notice that in the dimension $4<d<6$ the $\Phi^4$ operator is irrelevant as opposed to the $d<4$ case. 
  
  We pass the following parallels between the many-flavor bosonic ($4<d<6$) and many-flavor fermionic ($2<d<4$) QED's. The $U(N_f)$-Gross-Neveu-Yukawa model with $N_f$ four component fermions is the analog of $U(N_f)$-Yukawa model (this analogy was already pointed out in \cite{Fei:2014yja}). The $SU(N_f)$ pure fermionic  QED (no $\sigma\Psi^2$ interaction) is the analog of the $SU(N_f)$ pure scalar QED. The QED-GNY is the analog of the $\bC\bP^{(N_f-1)}$. This analogy lies on the following observation: the quartic interaction $\Phi^4$ is irrelevant in $4<d<6$, while the four fermion interaction $\Psi^4$ is irrelevant in $2<d<4$. The IR fixed points of the $U(N_f)$-Gross-Neveu-Yukawa and of the $SU(N_f)$ pure fermionic  QED in $2<d<4$ are respectively related to the UV fixed points of the Gross-Neveu and of the Thirring models \cite{Hasenfratz:1991it, ZinnJustin:1991yn}. See  \cite{Vasiliev:1992wr, Gracey:1993sn} for higher order in $1/N_f$ studies  for the fermionic models. 
  
\begin{figure}[H] 
 \centering
\includegraphics[width=0.8\textwidth]{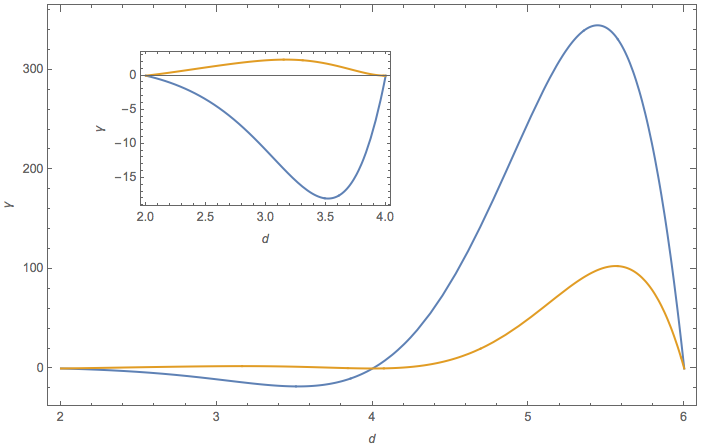} 
\caption{$\gamma_1(d)$ (blue) and $\gamma_2(d)$ (orange)} \label{mT7}
\end{figure}

  Let us review the findings of \cite{Vas:1983} about the critical $\bC\bP^{(N_f-1)}$ model in $d$-dimensions. The scaling dimension of the fundamental scalar field in the Landau gauge is 
  \begin{align}\label{vas4}
  &\Delta[\Phi_i]= \frac{d-2}{2} + \frac{1}{4}\Big( 1+\frac{4(d-1)^2}{d-4}\Big) \frac{\eta_1}{N_f}+ O\Big(\frac{1}{N_f^2}\Big) \ , \\
  & \text{where} \  \ \eta_1 \equiv  -\frac{4 a\big(2-\frac{d}{2} \big) a\big(\frac{d}{2}-1\big)}{a(2)\Gamma\big(\frac{d}{2}+1\big )} \text{ \ and \ } a(z)=\frac{\Gamma(d/2-z)}{\Gamma(z)} \ .
  \end{align}
 At the critical point, due to gauge invariance it is expected that the scaling dimension of the gauge field is exactly equal to $\Delta[A_\mu]=1$. The absence of the anomalous dimension was confirmed in \cite{Vas:1983} at the order $O(1/N_f)$. The scaling dimension of the HS field $\sigma$ is 
  \begin{align}\label{vas5}
  \Delta[\sigma]=2+\frac{d^2 (d-1)(2-d)}{4-d} \frac{\eta_1}{N_f}+ O\Big(\frac{1}{N_f^2}\Big) \ .
  \end{align}
  In \cite{Vasiliev:1975mq} it was observed, that at the critical point the condition $\Phi^2=0$ (which is equivalent to saying that the singlet quadratic operator is out of spectrum) doesn't hold after one introduces the analytic regularization. This regularization was employed in \cite{Vas:1983}. However in \cite{Vasiliev:1975mq} using the Schwinger-Dyson equations, it was proved that the $\Phi^2$ doesn't give any new scaling dimension, instead 
  \begin{align} \label{vas6}
  \Delta[\Phi^2]=d-\Delta[\sigma]=\nu^{-1} \ .
  \end{align}
This relation is known as a ``shadow relation".  The anomalous scaling dimensions of the operators $O_1=\frac{\sigma^2}{2}$ and $O_2=\frac{F_{\alpha\beta}^2}{4}$ was also studied. These operators have a scaling dimension 4 at leading order, and at the order $O(1/N_f)$ they mix. The mixing matrix in $d$-dimensions has the following form 
  \begin{align}
 \frac{\gamma}{N_f}= -\frac{4 a\big(2-\frac{d}{2} \big) a\big(\frac{d}{2}-1\big)}{N_f a(2)\Gamma\big(\frac{d}{2}+1\big )} \begin{bmatrix}
    \gamma_{11}       & \gamma_{12} \\
    \gamma_{21}       & \gamma_{22} 
\end{bmatrix} \ ,
  \end{align}
  where 
  \begin{align}
  &\gamma_{11}= \frac{d (d-1)^2(3-2d)}{4-d} \ , \\
  &\gamma_{12}=\frac{(4-d)(d+1)}{2} \ ,\\
  &\gamma_{21}=\frac{d(d-1)^3(d+1)}{4-d} \ , \\
  &\gamma_{22}=\frac{(d^2-d-4)(d-1)}{2} \ .
  \end{align} 
  The eigenvalues of the matrix $\gamma/N_f$ are the anomalous dimensions which we denote by $\gamma_{1,2}/N_f$. The eigenstates are mixtures of the operators $O_1$ and $O_2$. The full scaling dimensions are 
  \begin{align} \label{vas7}
  &\Delta_1=4+\frac{\gamma_1(d)}{N_f}+ O\Big(\frac{1}{N_f^2}\Big) \ , \\
&\Delta_2=4+\frac{\gamma_2(d)}{N_f}+ O\Big(\frac{1}{N_f^2}\Big) \ . \label{vas8}
  \end{align}
  The analytic expressions for $\gamma_{1,2}$ as a function of $d$ are very cumbersome. In Fig. \ref{mT7} we plot them in the region $2<d<6$ (and a separate small plot shows the same functions in the region $2<d<4$).
  
For our future purpose we will need the scaling dimensions (\ref{vas4}, \ref{vas5}, \ref{vas6}, \ref{vas7}, \ref{vas8}) in $d=6-2\epsilon$ expanded for small $\epsilon$
\begin{align} \label{vas9}
&\Delta[\Phi_i]=2-\epsilon+\frac{51}{N_f}\epsilon-\frac{167}{2N_f}\epsilon^2 +O(\epsilon^3) \ ,\\ \label{vas10}
&\Delta[\sigma]=2+\frac{1440}{N_f}\epsilon-\frac{3456}{N_f} \epsilon^2+O(\epsilon^3) \ , \\ \label{vas11} 
&\Delta[A_\mu]=1 \ , \\ \label{vas19}
&\Delta[\Phi^2]= 4-2\epsilon-\frac{1440}{N_f}\epsilon+\frac{3456}{N_f} \epsilon^2+O(\epsilon^3) \ , \\ \label{vas12}
&\Delta_1=4+\frac{40(50+7\sqrt{10})}{N_f}\epsilon-\frac{2(8275+827\sqrt{10})}{3N_f}\epsilon^2+O(\epsilon^3) \ ,\\ \label{vas13}
&\Delta_2=4+\frac{40(50-7\sqrt{10})}{N_f}\epsilon-\frac{2(8275-827\sqrt{10})}{3N_f}\epsilon^2+O(\epsilon^3) \ .
\end{align}
  To our knowledge the critical scalar-QED in $4<d<6$ has not been studied yet. We calculated scaling dimensions at order $O(1/N_f)$ for few operators. Below we give the results without providing details of the calculations 
  \begin{align} \label{vas14}
  &\Delta[\Phi_i]=\frac{d-2}{2}+\frac{(d-1)^2}{d-4} \frac{\eta_1}{N_f}+ O\Big(\frac{1}{N_f^2}\Big)  \ , \\ \label{vas15}
  &\Delta[A_\mu]=1 \ , \\ \label{vas16}
  &\Delta[\Phi^2]=d-2+\frac{(d-1)^2 (d(d-1)-2)}{4-d} \frac{\eta_1}{N_f}+ O\Big(\frac{1}{N_f^2}\Big) \ .
  \end{align}  
  The dimension of $\Phi_i$ (\ref{vas14}) is given in the Landau gauge. Expanding (\ref{vas14}, \ref{vas16}) at $d=6-2\epsilon$ for small $\epsilon$ gives 
  \begin{align} \label{vas17}
  &\Delta[\Phi_i]=2-\epsilon+\frac{50}{N_f}\epsilon-\frac{245}{3N_f}\epsilon^2+O(\epsilon^3)  \ , \\ \label{vas18}
  &\Delta[\Phi^2]=4-2\epsilon -\frac{1400}{N_f}\epsilon +\frac{10160}{3N_f}\epsilon^2+ O(\epsilon^3) \ .
  \end{align}

\section{Higher Derivative Gauge theory in $d=6$} \label{HDG}
The HDG is defined with the following Euclidean (bare) action 
\begin{align} \label{hard}
S=&\int d^d x \Big[\overline{D_\mu \Phi_i} D^\mu \Phi^i+\frac{1}{2} \partial_\mu \sigma \partial^\mu \sigma+\frac{1}{4}\partial_\mu F_{\alpha\beta} \partial^\mu F^{\alpha\beta}+
\tau_1^{(0)} \Phi^*_i \Phi^i+\frac{\tau_2^{(0)} \sigma^2}{2}+ \frac{\tau^{(0)}_3 F_{\alpha\beta}F^{\alpha\beta}}{4} \nonumber \\
&+g_1^{(0)} \sigma \Phi^*_i \Phi^i + \frac{ g_2^{(0)} \sigma^3}{6}+\frac{\lambda^{(0)}\sigma F_{\alpha\beta}F^{\alpha\beta} }{2}+
\frac{1}{2\xi} \big(\partial_\mu\partial_\alpha A^\alpha\big) \big(\partial^\mu \partial_\beta A^\beta\big)\Big] \ ,
\end{align} 
where $D_\mu = \partial_\mu +i e_0 A_\mu$. The action (\ref{hard}) has a $SU(N_f)$ global symmetry, the complex scalar fields $\Phi_i, \ i=1,...,N_f$ transform in the fundamental representation of $SU(N_f)$. The real scalar field $\sigma$ is a $SU(N_f)$ singlet. The kinetic term for the gauge field $A_\mu$ contains 4-derivatives as opposed to the standard two-derivative kinetic terms, hence the name ``higher derivative gauge theory". The last term in the action (\ref{hard}) is the gauge fixing term. We call it a $\mathbbm{R}_\xi$ gauge borrowing the name of the standard gauge fixing: $\frac{(\partial A)^2}{2\xi}$ , commonly used in the 4-dimensional gauge theories.
The propagator of the gauge field $A_\mu$ in the $\mathbbm{R}_\xi$ gauge has the following form. 
\begin{align} \label{sixdimphoton}
D_{\alpha\beta}(p)=\langle A_\alpha(p) A_\beta (-p) \rangle= \frac{1}{p^2 (p^2+\tau_3)} \Big[ \delta_{\alpha\beta} + \frac{(\xi-1)p^2+\xi \tau_3}{p^2}\frac{p_\alpha p_\beta}{p^2}\Big] \ .
\end{align}
We observe that in the 6-dimensional fermionic QED and fermionic QCD \cite{Gracey:2015xmw} the gauge field has the same propagator.  We will work in the Landau gauge $\xi=0$. In the Landau gauge the propagator is transverse: $D_{\alpha\beta}(p) p^\beta=0$.  

 The canonical dimensions of the scalar and gauge fields in $d=6$ are: $d[\Phi]=d[\sigma]=2, \ d[A]=1$. Following the general rules, in the action (\ref{hard}) we included all the possible terms (scalar gauge invariant operators preserving the $SU(N_f)$ symmetry) that have dimensions less or equal to 6. There are 3 mass terms: $\Phi^2, \sigma^2, F_{\alpha\beta}^2$ with dimensions equal to 4 (relevant operators) and there are 3 cubic interactions: $\sigma\Phi^2, \sigma^3, \sigma F_{\alpha\beta}^2$ with dimensions equal to 6 (marginal operators). The scalars $\Phi_i$ are minimally coupled to the gauge field which introduces the standard cubic and quartic interactions between these fields. To distinguish the bare parameters from the physical ones, we denoted the former with a superscript (\ref{hard}). 
 
 The marginal operator $F_{\alpha\beta} F_{\beta \gamma}F_{\gamma\alpha}$ is identically vanishing, since under the exchange $\alpha \leftrightarrow \beta$  the $F_{\alpha\beta}$ is antisymmetric and the $F_{\beta \gamma}F_{\gamma\alpha}$ is symmetric. Notice that besides the kinetic term for the gauge field that appears in (\ref{hard}), there is another dimension 6, 4-derivative operator: $\sim \partial_\mu F_{\alpha\beta} \partial_{\alpha} F_{\mu\beta}$. However we can prove that it is not an independent operator, indeed
\begin{align}
\partial_\mu F_{\alpha\beta} \partial_{\alpha} F_{\mu\beta}=\partial_\mu F_{\alpha\beta}  \big( \partial_{\alpha} F_{\mu\beta} - \partial_{\beta} F_{\mu\alpha}  \big)= \partial_\mu F_{\alpha\beta} \partial_\mu F_{\alpha\beta}  \ ,
\end{align}
where in the last step we used the Bianchi identity: $\partial_\mu F_{\alpha\beta} +\partial_\alpha F_{\beta\mu}+ \partial_\beta F_{\mu\alpha}=0$. 
Therefore we conclude that in the action (\ref{hard}) we should include only one of these 4-derivative operators, which is what we did.

In order to cure the divergencies appearing in the Green functions we need to renormalize the action (\ref{hard}). We perform the renormalization in the Minimal Subtraction (MS) scheme. First we introduce dimensional regularization, i.e. we define the theory (\ref{hard}) in the dimension $d=6-2\epsilon$. The canonical dimensions of the fields in $d=6-2\epsilon$ are: $d[\Phi]=d[\sigma]=2-\epsilon, \ d[A]=1-\epsilon$. The bare action (\ref{hard}) is related to the renormalized action by field renormalizations:
\begin{align}
S_R(\Phi,\sigma,A)&=S(Z_\Phi \Phi,Z_\sigma \sigma,Z_A A) \\
Z_\Phi&=Z_\Phi(g_1,g_2,e,\lambda,\epsilon) \label{m1} \\
Z_\sigma&=Z_\sigma(g_1,g_2,e,\lambda,\epsilon) \label{m2} \\
Z_A&=Z_A(g_1,g_2,e,\lambda,\epsilon) \label{m3} \ .
\end{align}
The bare masses $\tau^{(0)}_a$ are related to the renormalized masses $\tau_a$ 
\begin{align}
\tau^{(0)}_a=\sum_{b} Z_{ab}^\tau(g_1,g_2,e,\lambda,\epsilon)  \tau_b \ , \  \ a,b=1,2,3 \  \ .\label{m4}
\end{align}
The canonical dimensions of the mass parameters are $d[\tau_a^{(0)}]=d[\tau_a]=2$. The canonical dimensions of the bare couplings are $d[g_1^{(0)}]=d[g_2^{(0)}]=d[e^{(0)}]=d[\lambda^{(0)}]=\epsilon$. The renormalized couplings in (\ref{renorm_marginal}) are dimensionless, this is achieved by introducing the MS scheme parameter $\mu$, which has a mass dimension equal to one. For convenience let us denote $(e=g_3, \lambda=g_4)$, then the relation between the bare and renormalized couplings can be written in the compact form 
\begin{align} \label{renorm_marginal}
g^{(0)}_u=\sum_{v}Z_{u v} (g_1,g_2,e,\lambda,\epsilon) \mu^\epsilon g_v \ , \ \ u,v=1,2,3,4  \  \ .
\end{align}
The gauge coupling is actually renormalized multiplicatively  
\begin{align}
e^{(0)}=Z_e(g_1,g_2,e,\lambda,\epsilon) \mu^\epsilon e \label{m6} \ .
\end{align}
In other words $Z_{31}=Z_{32}=Z_{34}=0$ and $Z_{33} \equiv Z_e$ in (\ref{renorm_marginal}). It follows from the gauge invariance of the action (\ref{hard}) that $Z_e Z_A=1$. Therefore we do not need to separately renormalize the gauge interaction vertices ($A_\alpha^2 |\Phi|^2, A_\alpha \Phi^* \overset{\leftrightarrow}{\partial}_\alpha \Phi$), instead we determine $Z_A=1/Z_e$  by studing the renormalization of the gauge field propagator.

We remind that the renormalized action is a function either of bare parameters or of renormalized parameters, since only one set can be considered to be independent. We choose $S_R$ to be a function of renormalized masses and couplings. 
\begin{align} \label{renorm_action}
S_R= &\int d^d x \Big[\overline{D_\mu \Phi_i} D^\mu \Phi^i+\frac{1}{2} \partial_\mu \sigma \partial^\mu \sigma+\frac{1}{4}\partial_\mu F_{\alpha\beta} \partial^\mu F^{\alpha\beta}+
\tau_1 \Phi^*_i \Phi^i+\frac{\tau_2 \sigma^2}{2}+ \frac{\tau_3 F_{\alpha\beta}F^{\alpha\beta}}{4} \nonumber \\
&+g_1 \mu^\epsilon\sigma \Phi^*_i \Phi^i + \frac{ g_2 \mu^\epsilon \sigma^3}{6}+\frac{\lambda\mu^\epsilon\sigma F_{\alpha\beta}F^{\alpha\beta} }{2}+
\frac{1}{2\xi} \big(\partial_\mu\partial_\alpha A^\alpha\big) \big(\partial^\mu \partial_\beta A^\beta\big)  \nonumber \\
&+(Z_\Phi^2-1)\overline{D_\mu \Phi_i} D^\mu \Phi^i+\frac{Z_\sigma^2-1}{2} \partial_\mu \sigma \partial^\mu \sigma+\frac{Z^2_A-1}{4}\partial_\mu F_{\alpha\beta} \partial^\mu F^{\alpha\beta} \nonumber \\
&+ \big( Z_\Phi^2  \sum Z^\tau_{1a} \tau_a -\tau_1\big) \Phi^*_i \Phi^i+\frac{\big(  Z^2_\sigma \sum Z^\tau_{2a} \tau_a -\tau_2\big)
\sigma^2}{2}+\frac{\big(  Z^2_A \sum Z^\tau_{3a} \tau_a -\tau_3\big) F_{\alpha\beta}F^{\alpha\beta}}{4} \nonumber \\
&+\! \big( Z_\Phi^2 Z_\sigma\!  \sum\! Z_{1u} g_u -g_1\big) \mu^\epsilon \sigma \Phi^*_i \Phi^i+\frac{\big(  Z^3_\sigma \! \sum \! Z_{2u} g_u -g_2\big)\mu^\epsilon \sigma^3}{6} + \! \frac{\big( Z_\sigma Z_A^2 \! \sum Z_{4u} g_u - \! \lambda \big) \mu^\epsilon \sigma F_{\alpha\beta}F^{\alpha\beta}}{2}\! \Big] \ .
\end{align}
Using (\ref{renorm_action}) we define the Feynman rules for the vertices and for the counter-vertices (CV). The graphical representation for the propagators and vertices are collected in Tab. \ref{mT1}. 
\begin{align} \label{m22}
&CV^{(\sigma\Phi\Phi^*)} \!= \!-(Z_\Phi^2 Z_\sigma Z_{11}-1\!)g_1 \mu^\epsilon-\! Z_\Phi^2 Z_\sigma Z_{12}g_2 \mu^\epsilon\!-\! Z_\Phi^2 Z_\sigma Z_{13}e \mu^\epsilon\!-\!Z_\Phi^2 Z_\sigma Z_{14}\lambda \mu^\epsilon \ , \\ \label{m23} 
&CV^{(\sigma\sigma\sigma)}= - Z_\sigma^3 Z_{21} g_1 \mu^\epsilon-(Z_\sigma^3 Z_{22}-1) g_2 \mu^\epsilon-Z_\sigma^3 Z_{23}e \mu^\epsilon- Z_\sigma^3 Z_{24}\lambda \mu^\epsilon  \ , \\ \label{m24}
&CV^{\big(\sigma A_\alpha(p) A_\beta(q)\big)}\!= \! 2 \big[  Z^2_AZ_\sigma Z_{41} g_1\mu^\epsilon+\! Z^2_A Z_\sigma Z_{42} g_2\mu^\epsilon+Z^2_A Z_\sigma Z_{43} e\mu^\epsilon+(Z^2_A Z_\sigma Z_{44}-1) \lambda \mu^\epsilon   \!  \big] L_{\alpha\beta}(p,q) \ .
\end{align}
Where we defined $L_{\alpha\beta}(p,q) \equiv \delta_{\alpha\beta} p \cdot q -p_\beta q_\alpha \ $. The counter-terms for the kinetic and for the mass terms are given in the first lines of Tab. (\ref{mT2}, \ref{mT4}, \ref{mT5}, \ref{mT6}). 

The 1-PI Green-functions of the renormalized theory are constructed in the form of perturbative expansion in the renormalized couplings. All the terms in this expansion can be represented graphically: connected Feynman graphs with amputated external legs and such that cutting any single internal leg doesn't split the graph into disconnected components. The Feynman graphs already at one-loop typically are divergent integrals (when we put $\epsilon=0$). Demanding that the Green functions are free of divergencies one determines order-by-order the renormalization constants (Z's) defined in (\ref{m1}, \ref{m2}, \ref{m3}, \ref{m4}, \ref{renorm_marginal}, \ref{m6}) and  the counter-vertices.  In the next section we determine $Z_\Phi,Z_\sigma,Z_A$ and the the matrix $Z_{uv}$ (\ref{renorm_marginal}). To determine these constants it is sufficient to renormalize the 2-point and the 3-point Green functions in the massless limit: $\tau_a=0, \ a=1,2,3$.

  \begin{table}[]
  \centering 
  \begin{tikzpicture} [scale=0.7]  
  
  \draw[] (-3,2) to (13,2);
  \draw[black] (-3,2) to (-3,-20.5);
  \draw[black] (13,2) to (13,-20.5);
  \draw[black] (-3,-20.5) to (13,-20.5);
  
  \draw[thick, postaction={decorate}, decoration={markings, mark=at position .55 with {\arrow[scale=0.8]{triangle 45}}}] (-1.7,0.3) to (1.7,0.3); 
 \node at (7,0.3) {$=\langle \Phi_i(p) \Phi^*_j(-p)\rangle=\frac{\delta_{ij}}{p^2+\tau_1}$};
 \draw[dashed,thick] (-1.7,-1.3) to (1.7,-1.3);
 \node at (6.7,-1.3) {$=\langle \sigma(p) \sigma(-p)\rangle=\frac{1}{p^2+\tau_2}$};
 \draw[photon] (-1.7,-2.9) to (1.7,-2.9);
 \node at (7,-2.9) {$=\langle A_\alpha(p) A_\beta (-p) \rangle=(\ref{sixdimphoton})$};

 \begin{scope}[shift={(0,-4.1)}]
 \draw[dashed,thick] (0,-0.2) to (0,-1.5);
 \draw[thick, postaction={decorate}, decoration={markings, mark=at position .6 with {\arrow[scale=0.8]{triangle 45}}}] (-1,-2.5) to (0,-1.5);
 \draw[thick,postaction={decorate}, decoration={markings, mark=at position .31 with {\arrowreversed[scale=0.8]{triangle 45}}}] (1,-2.5) to (0,-1.5);
 \node at (5,-1.5) {$=-\mu^\epsilon g_1$};
 \end{scope}
 
 \begin{scope}[shift={(0,-7.2)}]
 \draw[dashed,thick] (0,-0.2) to (0,-1.5);
 \draw[thick,dashed] (-1,-2.5) to (0,-1.5);
 \draw[thick,dashed] (1,-2.5) to (0,-1.5);
  \node at (5,-1.5) {$=-\mu^\epsilon g_2$};
 \end{scope}
 
  \begin{scope}[shift={(0,-10.4)}]
 \draw[photon1] (0,-0.2) to (0,-1.5);
 \draw[thick, postaction={decorate}, decoration={markings, mark=at position .6 with {\arrow[scale=0.8]{triangle 45}}}] (-1,-2.5) to (0,-1.5);
 \node at (-1,-1.8) {$q$};
 \draw[thick,postaction={decorate}, decoration={markings, mark=at position .31 with {\arrowreversed[scale=0.8]{triangle 45}}}] (1,-2.5) to (0,-1.5);
  \node at (1,-1.8) {$p$};
  \node at (6,-1.5) {$=\mu^\epsilon e(p+q)_\alpha$};
 \end{scope}
 
 \begin{scope}[shift={(0,-13.8)}]
 \draw[photon1] (-0.8,-0.3) to (0,-1.5);
  \draw[photon1] (0.8,-0.25) to (0,-1.5);
 \draw[thick, postaction={decorate}, decoration={markings, mark=at position .6 with {\arrow[scale=0.8]{triangle 45}}}] (-1,-2.5) to (0,-1.5);
 \draw[thick,postaction={decorate}, decoration={markings, mark=at position .31 with {\arrowreversed[scale=0.8]{triangle 45}}}] (1,-2.5) to (0,-1.5);
  \node at (6,-1.5) {$=-2\mu^{2\epsilon}\delta_{\alpha\beta}e^2$};
 \end{scope}
 
  \begin{scope}[shift={(0,-16.9)}]
 \draw[dashed,thick] (0,-0.2) to (0,-1.5);
 \draw[photon1] (-1,-2.5) to (0,-1.5);
  \draw[ <-] (-1.5+0.5,-2.5+0.5) to (0-0.4,-1-0.4);
  \node at (-1,-1.5 ) {$p$};
  \draw[->] (0.1+0.4-0.1,-1-0.4) to (2-0.9-0.1,-3+1);
    \node at (1,-1.5 ) {$q$};
 \draw[photon1] (1,-2.5) to (0,-1.5);
\node at (-1.2,-2.7) {$\alpha$}; 
\node at (1.2,-2.7) {$\beta$}; 
  \node at (7.5,-1.7) {$=2\mu^\epsilon \lambda (\delta_{\alpha\beta} p\cdot q-p_\beta q_\alpha)=2\mu^\epsilon \lambda L_{\alpha\beta}(p,q)$};
 \end{scope}

 \end{tikzpicture}
\caption{ Feynman rules for tree-level propagators and vertices}  \label{mT1}
 \end{table}

 \section{Renormalization of fields and of cubic vertices: anomalous dimensions of fields and beta functions }
 
 We study the 1-PI 2-point Green-functions for the scalar and gauge fields at the one-loop order. The Tab. \ref{mT2} contain all the one-loop graphs that appear in those Green-functions. For our purposes, it is sufficient to calculate the divergent parts of the one-loop integrals, which are (simple) poles in $\epsilon\rightarrow 0$. Some of the graphs ($G_1, G_4, G_5$) in Tab. \ref{mT2} have already been evaluated in the context of the $O(N)$-Yukawa theory \cite{Fei:2014yja}, which is the ungauged version of our theory\footnote{ More precisely one should take $N=2N_f$ in the $O(N)$-Yukawa theory, then to gauge the $U(1)$ factor in the $U(1)\times SU(N_f)\subset O(2N_f)$. As a result one will obtain the $SU(N_f)$ symmetric higher derivative gauge theory (\ref{hard}).}.
 
 Using (\ref{ma2}), we obtain for the graph $G_1$ 
 \begin{align} \nonumber
 G_1&= (-g_1)^2 \mu^{2\epsilon} \int \frac{d^d q}{(2\pi)^d} \frac{1}{q^2(q+p)^2}= (-g_1)^2 \mu^{2\epsilon} \frac{\Gamma(2-\epsilon)^2 \Gamma(-1+\epsilon)}{(4\pi)^{3-\epsilon} \Gamma(1)^2 \Gamma(4-2\epsilon)} p^{2-2\epsilon}\\
 &\overset{\epsilon\rightarrow 0}{=}-\frac{g_1^2}{6(4\pi)^3\epsilon} p^2 \label{m7} \ .
 \end{align}
  The graph $G_2$ gives
 \begin{align} \nonumber 
 G_2&=e^2 \mu^{2\epsilon} \int \frac{d^d q}{(2\pi)^d} \frac{(2p+q)_\alpha (2p+q)_\beta} {(p+q)^2} \frac{\delta_{\alpha\beta}-\frac{q_\alpha q_\beta}{q^2}}{q^4} \\
 &=4e^2 \mu^{2\epsilon} \Big(  p^2 \int \frac{d^d q}{(2\pi)^d} \frac{1}{(p+q)^2 q^4}- p_\alpha p_\beta  \int \frac{d^d q}{(2\pi)^d} \frac{q_\alpha q_\beta}{(p+q)^2 q^6}\Big) \nonumber \\
 &\overset{\epsilon\rightarrow 0}{=} 4e^2 \Big( \frac{p^2}{2(4\pi)^3 \epsilon}- \frac{p^2}{12(4\pi)^3 \epsilon} \Big)= \frac{5e^2}{3(4\pi)^3 \epsilon} p^2  \label{m8} \ .
 \end{align}
 To pass to the second line in (\ref{m8}), we used the transversality condition of the photon propagator. The first integral of the second line is evaluated using (\ref{ma2}), the second integral is evaluated introducing Feynman parametrization (\ref{ma3}) and then using formulas (\ref{ma1}, \ref{ma4}). The tadpole $G_3$ is vanishing in the dimensional regularization in the massless limit and therefore it does not contribute to the field renormalization. However we will see in the next section that the tadpoles are important for mass renormalizations.  The counter-term $CV^{(\Phi\Phi^*)}= -(Z_\Phi^2-1)p^2$ must be such that the Green function $\Gamma^{(\Phi\Phi^*)}$ is finite, thus
 \begin{align} \label{m19}
 Z_\Phi=1-\frac{g_1^2}{12(4\pi)^3 \epsilon}+\frac{5e^2}{6(4\pi)^3 \epsilon} \ .
 \end{align}
 The graphs $G_4, G_5$ have the same topology as the graph $G_1$, and can be evaluated similarly. Their values are given in Tab. \ref{mT2} .The graph $G_6$ gives
 \begin{align} \label{m9}
 G_6=2\lambda^2 \mu^{2\epsilon} \int \frac{d^d q}{(2\pi)^d} L_{\alpha\mu}(q,p-q) D_{\alpha\beta}(q) L_{\beta\nu}(-q,q-p) D_{\mu\nu}(p-q)\overset{\epsilon\rightarrow 0}{=} -\frac{5\lambda^2}{(4\pi)^3\epsilon} p^2 \ .
 \end{align}
  The integral in (\ref{m9}) can be simplified using the transversality condition and the identity $q(q-p)=\frac{q^2+(q-p)^2-p^2}{2}$. The resulting integrals are evaluated introducing the Feynman parametrization and with the help of formulas (\ref{ma1}, \ref{ma4}, \ref{ma5}). We omit the details of a long and tedious calculation. 
  
  The counter-term $CV^{(\sigma\sigma)}=-(Z_\sigma^2-1)p^2$ should cancel the divergencies in the Green function $\Gamma^{(\sigma \sigma)}$, thus
  \begin{align} \label{m20}
  Z_\sigma= 1- \frac{N_f g_1^2}{12(4\pi)^3 \epsilon}-\frac{g_2^2}{24(4\pi)^3\epsilon}-\frac{5\lambda^2}{2(4\pi)^3\epsilon} \ .
  \end{align}
  The graph $G_7$ gives
  \begin{align}  \label{m11}
G_7=N_f e^2 \mu^{2\epsilon}  \int \frac{d^d q}{(2\pi)^d} \frac{(p+2q)_\alpha (p+2q)_\beta} {q^2 (p+q)^2}\overset{\epsilon\rightarrow 0}{=} \frac{N_f e^2}{30(4\pi)^3\epsilon} (\delta_{\alpha\beta}p^4-p_\alpha p_\beta p^2) \ .
 \end{align}
 The factor $N_f$ is due to the $N_f$ scalar flavors circulating in the loop of the graph $G_7$. Notice that the $G_7$ (\ref{m11}) is transverse. This was expected since $G_7$ contributes to the self-energy of the photon, which in turn must be transverse due to the gauge invariance. The tadpole $G_8$ is vanishing in the dimensional regularization in the massless limit. The graph $G_9$ has no pole
 \begin{align}  \label{m12}
G_9= 4\lambda^2 \mu^{2\epsilon}  \int \frac{d^d q}{(2\pi)^d} \frac{1}{(p-q)^2} L_{\alpha\mu}(-p,q) D_{\mu\nu}(q) L_{\nu\beta}(-q,p) \overset{\epsilon\rightarrow 0}{=} 0 \ .
\end{align}
The integral (\ref{m12}) is simplified noticing that $D_{\mu\nu}(q) L_{\nu\beta}(-q,p)=L_{\mu\beta}/q^4$, the resulting integral is easily calculated with the help of formulas of appendix \ref{usefulformulas}.

We choose the $Z_A$ such that the counter-term $CV^{(AA)}=-(Z^2_A-1)(\delta_{\alpha\beta}p^4-p_\alpha p_\beta p^2)$ cancels the divergencies in the Green function $\Gamma^{(AA)}$ 
\begin{align} \label{m21}
Z_A=1+\frac{N_f e^2}{60(4\pi)^3 \epsilon} \ .
\end{align}
The anomalous dimensions of the fields are constructed using the field renormalization constants (\ref{m19}, \ref{m20}, \ref{m21}) as follows 
\begin{align} \label{m25}
&\gamma_\Phi=\frac{d\ln Z_\Phi}{d \ln\mu}= \frac{g_1^2-10e^2}{6(4\pi)^3} \ , \\ \label{m26}
&\gamma_\sigma=\frac{d\ln Z_\sigma}{d \ln\mu}= \frac{2 N_f g_1^2+g_2^2+60\lambda^2}{12(4\pi)^3}  \ ,\\
&\gamma_A=\frac{d\ln Z_A}{d \ln\mu}= -\frac{N_f e^2}{30(4\pi)^3} \label{m27} \ ,
\end{align}
where we used the chain rule $\frac{d\ln Z}{d \ln\mu}=\sum\limits_{u} \beta_{g_u} \frac{d\ln Z}{d g_u}$ and the beta functions in the trivial (classical) approximation $\beta_{g_u}=(-\epsilon g_u+...)$ .

 \begin{table}[]
 \centering 
 \begin{tikzpicture} [scale=0.8]  
  
  \draw[] (-1,4) to (19,4);
  \draw[] (-1,4) to (-1,-4.8);
    \draw[] (-1,-4.8) to (19,-4.8);
    \draw[] (9,4) to (9,-4.8);
     \draw[] (19,4) to (19,-4.8);
    \draw[] (-1,3) to (19,3);
      
   \node at (3.5,3.5) {$\Gamma^{(\Phi\Phi^*)}$};            
    \node at (13.5,3.5) {$\Gamma^{(\sigma \sigma)}$};

 \draw[thick] (0,2) to (3,2);
 \node at (6.5,2) {$= \ \ -(Z_\Phi^2-1)p^2$};
\draw[fill] (1.5,2) circle (4pt);   
\draw[thick] (0,0) to (3,0);
\draw[dashed, thick] ([shift=(180:1cm)]1.5,0) arc (180:0:1 cm);
\node at (6.2,0) {$= \ \ -\frac{g_1^2 }{6(4\pi)^3 \epsilon} p^2$};
\node at (-0.5,0) {$\scriptstyle{G_1}$};
\draw[thick] (0,-2) to (3,-2);
\draw [wavy semicircle, thick] (0.5,-2 ) to  (2.5,-2);
\node at (6.2,-2) {$= \ \ -\frac{5e^2 }{3(4\pi)^3 \epsilon} p^2$};
\node at (-0.5,-2) {$\scriptstyle{G_2}$};
 \draw[thick] (0,-4) to (3,-4);
 \draw[thick,decorate,decoration={complete siness}] (0.9,-3.26) arc (180-5:-180-5:0.6);
 \node at (-0.5,-4) {$\scriptstyle{G_3}$};
 \node at (5.2,-4) {$=\ 0$};

  \begin{scope}[shift={(10,0)}]
  \draw[thick,dashed] (0,2) to (3,2);
 \node at (6.2,2) {$= \ \ -(Z_\sigma^2-1)p^2$};
\draw[fill] (1.5,2) circle (4pt); 
\node at (-0.5,0) {$\scriptstyle{G_4}$};
\node at (1.5,0) {$\scriptstyle{N_f}$};
\draw[dashed,thick] (0,0) to (0.8,0);
\draw[thick] ([shift=(180:1cm)]1.8,0) arc (180:-180:0.7 cm);
\draw[dashed,thick] (2.2,0) to (3.03,0);
\node at (-0.5,-2.2) {$\scriptstyle{G_5}$};
\node at (6,0) {$= \ \ -\frac{N_f g_1^2 }{6(4\pi)^3 \epsilon} p^2$};
\draw[dashed,thick] (0,-2.2) to (0.8,-2.2);
\draw[dashed, thick] ([shift=(180:0.7cm)]1.5,-2.2) arc (180:-180:0.7 cm);
\node at (6.2,-2.2) {$= \ \ -\frac{g_2^2 }{12(4\pi)^3 \epsilon} p^2$};
\node at (-0.5,-4) {$\scriptstyle{G_6}$};
\draw[dashed,thick] (2.2,-2.2) to (3.03,-2.2);
\begin{scope}[shift={(0,-0.2)}]
\draw[dashed,thick] (0,-3.8) to (0.9,-3.8);
 \draw[thick,decorate,decoration={complete siness}] (0.9,-3.65) arc (180-5:-180-5:0.6);
 \draw[dashed,thick] (2.2,-3.8) to (3.03,-3.8);
 \node at (5.9,-3.8) {$=\ -\frac{5\lambda^2}{(4\pi)^3\epsilon} p^2$};
\end{scope}

 \end{scope}
  
   \begin{scope}[shift={(4.5,-8)}]
   \draw[] (-1,2) to (12,2);
    \draw[] (-1,2) to (-1,-8);
    \draw[] (12,2) to (12,-8);
    \draw[] (-1,-8) to (12,-8);
  \draw[] (-1,1) to (12,1);
  
  \node at (5.5,1.5) {$\Gamma^{(AA)}$};

   \draw[photon2] (0,0) to (3.4,0);
   \draw[fill] (1.7,0) circle (4pt);   
   \node at (8.3,0) {$=\ \ -(Z^2_A-1)(\delta_{\alpha\beta}p^4-p_\alpha p_\beta p^2)$};
   \node at (-0.5,-2) {$\scriptstyle{G_7}$};
    \draw[photon3] (0,-2) to (1,-2);
    \draw[ thick] ([shift=(180:1cm)]2,-2) arc (180:-180:0.7 cm);
    \draw[photon3] (2.4,-2) to (3.4,-2);
   \node at (7.9,-2) {$= \ \ \frac{N_f e^2}{30(4\pi)^3\epsilon}(\delta_{\alpha\beta}p^4-p_\alpha p_\beta p^2)$};
    \node at (1.7,-2) {$\scriptstyle{N_f}$};
      \node at (-0.5,-4.7) {$\scriptstyle{G_8}$};
      \node at (5.4,-4.7) {$ = \ 0$};
            \node at (1.8,-4.2) {$\scriptstyle{N_f}$};
    \draw[photon2] (0,-4.8) to (1.7,-4.8);
    \draw[photon2] (1.7,-4.8) to (3.4,-4.8);
     \draw[thick] (1.75,-4.13) circle (0.6cm); 
    \draw[photon2] (0,-4.4-2.5) to (3.4,-4.4-2.5);
   \draw[dashed, thick] ([shift=(180:0.9cm)]1.8,-4.3-2.5) arc (180:0:0.9 cm);
     \node at (-0.5,-4.4-2.5) {$\scriptstyle{G_9}$};
   \node at (5.5,-4.4-2.5) {$=\ 0$};
   \end{scope}
  
 \end{tikzpicture}
\caption{ 2-point Green functions in the one-loop approximation.}  \label{mT2}
\end{table}

 \begin{table}[]
 \centering 
 \begin{tikzpicture} [scale=0.65]  
 
 \draw[] (-3,2.5+3) to (23.5,2.5+3);
 \draw[] (-3,1+3) to (23.5,1+3);
  \draw[] (-3,2.5+3) to (-3,-17);
    \draw[] (8.66-3,2.5+3) to (8.66-3,-17);
   \draw[] (17.33-3,2.5+3) to (17.33-3,-17);
  \draw[] (23.5,2.5+3) to (23.5,-17);
\draw[] (-3,-17) to (23.5,-17);

\node at (1,1.8+3) {$\Gamma^{(\sigma\Phi\Phi^*)}$};

\node at (10,1.8+3) {$\Gamma^{(\sigma\sigma\sigma)}$};

\node at (18.9,1.8+3) {$\Gamma^{(\sigma A A)}=\langle A_{\alpha}(p) A_\beta(-p) \sigma(0)\rangle$};

\node at (-2,-1) {$\scriptstyle{G_{10}}$};
\node at (-2,-4.5) {$\scriptstyle{G_{11}}$};
\node at (-2,-8) {$\scriptstyle{G_{12}}$};
\node at (-2,-11.5) {$\scriptstyle{G_{13}}$};
\node at (-2,-14.5) {$\scriptstyle{G_{14}}$};
\node at (6.5,-1) {$\scriptstyle{G_{15}}$};
\node at (6.5,-4.5) {$\scriptstyle{G_{16}}$};
\node at (6.5,-8) {$\scriptstyle{G_{17}}$};
\node at (15.1,-1) {$\scriptstyle{G_{18}}$};
\node at (15.1,-4.5) {$\scriptstyle{G_{19}}$};
\node at (15.1,-8) {$\scriptstyle{G_{20}}$};
\node at (15.1,-11.5) {$\scriptstyle{G_{21}}$};

\begin{scope}[shift={(-0.5,3)}]
 \draw[dashed,thick] (0,0) to (0,-1);
 \draw[ thick] (-1,-2) to (0,-1);
 \draw[ thick] (1,-2) to (0,-1);
   \draw[fill] (0,-1) circle (4pt);   
 \end{scope}

  \begin{scope}[shift={(-0.5,0)}]
 \draw[dashed,thick] (0,0) to (0,-1);
 \draw[ thick] (-1,-2) to (0,-1);
 \draw[ thick] (1,-2) to (0,-1);
 \draw[thick] (-2,-2) to (-1,-2);
  \draw[dashed,thick] (-1,-2) to (1,-2);
   \draw[thick] (1,-2) to (2,-2);
 \node at (4.5,-1.5) {$= \ -\frac{g_1^3 }{2(4\pi)^3 \epsilon}$};
 \end{scope}

   \begin{scope}[shift={(-0.5,-3.5)}]
 \draw[dashed,thick] (0,0) to (0,-1);
 \draw[dashed, thick] (-1,-2) to (0,-1);
 \draw[dashed, thick] (1,-2) to (0,-1);
 \draw[thick] (-2,-2) to (2,-2);
 \node at (4.5,-1.5) {$=\ -\frac{g_1^2 g_2}{2(4\pi)^3 \epsilon}$};
 \end{scope}
 
   \begin{scope}[shift={(-0.5,-7)}]
  \draw[dashed,thick] (0,0) to (0,-1);
 \draw[ thick] (-1,-2) to (0,-1);
 \draw[ thick] (1,-2) to (0,-1);
 \draw[thick] (-2,-2) to (-1,-2);
  \draw[photon3] (-1,-2) to (1,-2);
   \draw[thick] (1,-2) to (2,-2);
  \node at (3.7,-1.5) {$= \ 0$};
 \end{scope}
 
  \begin{scope}[shift={(-0.5,-10.5)}]
  \draw[dashed,thick] (0,0) to (0,-1);
 \draw[ photon3] (-1,-2) to (0,-1);
 \draw[ photon3] (1,-2) to (0,-1.1);
 \draw[thick] (-2,-2) to (-1,-2);
  \draw[thick] (-1,-2) to (1,-2);
   \draw[thick] (1,-2) to (2,-2);
  \node at (3.7,-1.5) {$= \ 0$};
 \end{scope}

  \begin{scope}[shift={(-2.3,-12)}]
  \draw[dashed,thick] (1.45,-1.6) to (1.45,-2.6);
 \draw[thick] (0,-4) to (3,-4);
 \draw[thick,decorate,decoration={complete siness}] (0.9,-3.26) arc (180-5:-180-5:0.6);
 \node at (6,-3.5) {$=  \frac{5e^2\lambda}{(4\pi)^3 \epsilon}$};
  \end{scope}

  \begin{scope}[shift={(-0.5+8.5,3)}]
 \draw[dashed,thick] (0,0) to (0,-1);
 \draw[dashed, thick] (-1,-2) to (0,-1);
 \draw[ dashed,thick] (1,-2) to (0,-1);
   \draw[fill] (0,-1) circle (4pt);   
 \end{scope}

  \begin{scope}[shift={(-0.5+8.5,0)}]
 \draw[dashed,thick] (0,0) to (0,-1);
 \draw[ thick] (-1,-2) to (0,-1);
 \draw[ thick] (1,-2) to (0,-1);
 \draw[dashed,thick] (-2,-2) to (-1,-2);
  \draw[thick] (-1,-2) to (1,-2);
   \draw[dashed,thick] (1,-2) to (2,-2);
 \node at (4.5,-1.5) {$= \ -\frac{N_f g_1^3 }{(4\pi)^3 \epsilon}$};
 \node at (0.07,-1.6) {$\scriptstyle{N_f}$};
 \end{scope}

   \begin{scope}[shift={(-0.5+8.5,-3.5)}]
 \draw[dashed,thick] (0,0) to (0,-1);
 \draw[dashed, thick] (-1,-2) to (0,-1);
 \draw[dashed, thick] (1,-2) to (0,-1);
 \draw[dashed,thick] (-2,-2) to (2,-2);
 \node at (4.5,-1.5) {$=\ -\frac{g_2^3 }{2(4\pi)^3 \epsilon}$};
 \end{scope}
 
   \begin{scope}[shift={(-0.5+8.5,-7)}]
  \draw[dashed,thick] (0,0) to (0,-1);
 \draw[ photon3] (-1,-2) to (0,-1);
 \draw[ photon3] (1,-2) to (0,-1);
 \draw[dashed,thick] (-2,-2) to (-1,-2);
  \draw[photon3] (-1,-2) to (1,-2);
   \draw[dashed,thick] (1.1,-2) to (2,-2);
  \node at (4.5,-1.5) {$= \ -\frac{20\lambda^3}{(4\pi)^3 \epsilon}$};
 \end{scope}

 \begin{scope}[shift={(-0.5+17,3)}]
 \draw[dashed,thick] (0,0) to (0,-1);
 \draw[ photon3] (-1,-2) to (0,-1);
 \draw[ photon3] (1,-2) to (0,-1);
   \draw[fill] (0,-1) circle (4pt);   
 \end{scope}

  \begin{scope}[shift={(-0.5+17,0)}]
 \draw[dashed,thick] (0,0) to (0,-1);
 \draw[ thick] (-1,-2) to (0,-1);
 \draw[ thick] (1,-2) to (0,-1);
 \draw[photon3] (-2,-2) to (-1,-2);
  \draw[thick] (-1,-2) to (1,-2);
   \draw[photon3] (1,-2) to (2,-2);
 \node at (4.5,-1.5) {$= \ -\frac{N_f g_1 e^2 }{3(4\pi)^3 \epsilon}L_{\alpha\beta}$};
 \node at (0.05,-1.6) {$\scriptstyle{N_f}$};
 \end{scope}

   \begin{scope}[shift={(-0.5+17,-3.5)}]
 \draw[dashed,thick] (0,0) to (0,-1);
 \draw[dashed, thick] (-1,-2) to (0,-1);
 \draw[dashed, thick] (1,-2) to (0,-1);
 \draw[photon3] (-2,-2) to (2,-2);
 \node at (4.5,-1.5) {$=\ \frac{2g_2 \lambda^2 }{3(4\pi)^3 \epsilon} L_{\alpha\beta}$};
 \end{scope}
 
   \begin{scope}[shift={(-0.5+17,-7)}]
  \draw[dashed,thick] (0,0) to (0,-1);
 \draw[ photon3] (-1,-2) to (0,-1);
 \draw[ photon3] (1,-2) to (0,-1);
 \draw[photon3] (-2,-2) to (-1,-2);
  \draw[thick,dashed] (-1+0.03,-2) to (1,-2);
   \draw[photon3] (1,-2) to (2,-2);
  \node at (4.5,-1.5) {$= \ \frac{4\lambda^3}{3(4\pi)^3\epsilon} L_{\alpha\beta}$};
 \end{scope}

 \begin{scope}[shift={(-2.3+17,-8.5)}]
  \draw[dashed,thick] (1.45,-1.6) to (1.45,-2.66);
 \draw[photon3] (0,-4) to (3,-4);
 \draw[thick] (0.9,-3.26) arc (180-5:-180-5:0.6);
 \node at (5.5,-3.5) {$= \ 0$};
 \node at (1.5,-3.3) {$\scriptstyle{N_f}$};
  \end{scope}

 \end{tikzpicture}
\caption{ 3-point Green functions in the one-loop approximation. $L_{\alpha\beta}=p_\alpha p_\beta-\delta_{\alpha\beta} p^2.$  }  \label{mT3}
\end{table}

  Next we proceed with the renormalization of the 3-point 1-PI Green functions, i.e. the vertices. All the one-loop graphs appearing in those Green functions are collected in Tab. \ref{mT3}. In the first line of Tab. \ref{mT3} we draw the counter-vertices. To calculate the graphs that appear in the Green functions $\Gamma^{(\sigma\Phi\Phi^*)}, \Gamma^{(\sigma\sigma\sigma)}$ we do not need to prescribe arbitrary momenta to the external legs (subject to the obvious condition that the total sum of the momenta is zero), it is sufficient to choose two of the external legs with opposite non-zero momenta and the remaining leg with zero momentum. Crucially the choice can vary from graph to graph, the only requirement is that the graph with a given choice of momenta ``leak" should not have infrared divergencies. This freedom, known as  ``Infrared rearrangement" \cite{Vladimirov:1979zm, Vasilbook}, is due to the graphs in the Green functions $\Gamma^{(\sigma\Phi\Phi^*)}, \Gamma^{(\sigma\sigma\sigma)}$ being only logarithmically divergent. With a good choice of momenta leak, the integrals can simplify a lot (this could be especially useful if one wants to do higher loop calculations). In the case of Green function $\Gamma^{(\sigma A A)}$, the graphs are quadratically divergent and so the momenta leak should be fixed and must be the same for all the graphs. 
  
    The graphs $G_{10}, G_{11}, G_{15}, G_{16}$ have already been evaluated in \cite{Fei:2014yja}. In all the one-loop graphs of $\Gamma^{(\sigma\Phi\Phi^*)}$, we choose the momenta leak in the external lines as follows: $\sigma(-p) \ \Phi(p) \  \Phi^*(0)$. The graph $G_{10}$ gives
  \begin{align}  \label{m13}
  G_{10}= (-g_1)^3 \mu^{3\epsilon}  \int \frac{d^d q}{(2\pi)^d} \frac{1}{(p+q)^2 q^4}\overset{\epsilon\rightarrow 0}{=} -\frac{g_1^3}{2(4\pi)^3\epsilon} \ .
  \end{align}
  The graph $G_{11}$ has the same topology as the graph $G_{10}$, and it can be evaluated similarly. Its value is given in Tab. \ref{mT3}. It is easy to see that $G_{12}, G_{13}$ have no poles. This is because in each of these graphs the right external leg carries a zero momentum (with our choice) and therefore the internal propagators adjacent to it  have the same momenta. Then it follows using the transversality condition that the integral vanishes. The graph $G_{14}$ gives 
 \begin{align} \nonumber  
  G_{14}=& -2e^2 \lambda \mu^{3\epsilon}  \int \frac{d^d q}{(2\pi)^d} \delta_{\alpha\beta} D_{\alpha\mu}(q) D_{\beta\nu}(p-q) L_{\mu\nu}(q,p-q)  \\ \nonumber
  =&\ 2e^2 \lambda\mu^{3\epsilon}  \int \frac{d^d q}{(2\pi)^d} \Big(\frac{d-2}{q^2(p-q)^4}+\frac{[q\cdot(p-q)]^2}{q^4(p-q)^6} \Big) \\
  \overset{\epsilon\rightarrow 0}{=} &\  2e^2\lambda \Big(\frac{2}{(4\pi)^3 \epsilon}+\frac{1}{2(4\pi)^3 \epsilon} \Big)=\frac{5e^2\lambda}{(4\pi)^3\epsilon} \label{m14} \ .
  \end{align}
   To pass to the second line we replaced $L_{\mu\nu}(q,p-q)=\delta_{\mu\nu}q\cdot (p-q)-q_\nu (p-q)_\mu \rightarrow -\delta_{\mu\nu}q^2$, because the terms in $L_{\mu\nu}(q,p-q)$ that are linear in $q$ will not give poles after integration and the term $q_\mu q_\nu$ gives zero contribution after tensor contraction with the photon propagators. In the counter-vertex $CV^{(\sigma\Phi\Phi^*)}$ (\ref{m22}) we choose the constants $(Z^2_\Phi Z_\sigma Z_{1u}, \ u=1,2,3,4)$  as to cancel the divergencies coming from $G_{10}, G_{11}, G_{14}$. Then using values of $Z_\Phi$ and $Z_\sigma$ from (\ref{m19}, \ref{m20}) we find the $( Z_{1u}, \ u=1,2,3,4)$
  \begin{align}
  &Z_{11}=1+\frac{(N_f-4) g_1^2}{12 (4\pi)^3\epsilon}-\frac{g_1 g_2}{2(4\pi)^3 \epsilon} +\frac{g_2^2}{24(4\pi)^3\epsilon}-\frac{5e^2}{3(4\pi)^3 \epsilon}+\frac{5\lambda^2}{2(4\pi)^3\epsilon}  \ , \\  
 & Z_{12}=Z_{13}=0 \ , \\  
 &Z_{14}=\frac{5 e^2}{(4\pi)^3\epsilon} \ .
  \end{align}
In all the one-loop graphs of $\Gamma^{(\sigma\sigma\sigma)}$ we choose the momenta leak as follows: $\sigma(-p) \ \sigma(p) \  \sigma(0)$. The graphs $G_{15}, G_{16}$ have the same topology as $G_{10}$, and can be evaluated similarly. Their values are reported in Tab. \ref{mT3}. The graph $G_{17}$ gives
 \begin{align} \nonumber 
G_{17} &= 8\lambda^3 \mu^{3\epsilon}  \int \frac{d^d q}{(2\pi)^d}  L_{\alpha\rho}(q,p-q) D_{\alpha\beta}(q) L_{\beta\mu}(-q,q) D_{\mu\nu}(q) L_{\nu\sigma}(-q,q-p) D_{\rho\sigma}(p-q) \\
&\overset{\epsilon\rightarrow 0}{=}-\frac{20 \lambda^3}{(4\pi)^3 \epsilon} \label{m15} \ .
\end{align}
 Demanding the counter-vertex $CV^{(\sigma\sigma\sigma)}$ (\ref{m23}) to cancel the divergencies coming from $G_{15}, G_{16}, G_{17}$ (i.e. to render the Green function $\Gamma^{(\sigma\sigma\sigma)}$ finite) and 
using the value of $Z_\sigma$ from (\ref{m20}) we find the $( Z_{2u}, \ u=1,2,3,4)$
\begin{align}
&Z_{21}= -\frac{N_f g_1^2}{(4\pi)^3\epsilon} \ , \\
&Z_{22}=1+\frac{N_f g_1^2}{4(4\pi)^3\epsilon}-\frac{3g_2^2}{8(4\pi)^3\epsilon}+\frac{15\lambda^2}{2(4\pi)^3\epsilon} \ , \\
&Z_{23}=0  \ ,\\
&Z_{24}=-\frac{20\lambda^2}{(4\pi)^3\epsilon} \ .
\end{align}
In the Green function $\Gamma^{(\sigma A A)}$ we choose the momenta leak as follows: $A_\alpha(p) \ A_\beta(-p) \ \sigma(0)$. The graphs $G_{18}, G_{19}, G_{20}$ give
\begin{align}
G_{18}=-2N_fg_1 e^2  \mu^{3\epsilon} \int \frac{d^d q}{(2\pi)^d} \frac{(p+2q)_\alpha (p+2q)_\beta}{q^4 (q+p)^2}\overset{\epsilon\rightarrow 0}{=} -\frac{N_f g_1 e^2}{3(4\pi)^3 \epsilon} L_{\alpha\beta} \label{m16} \ ,
\end{align}
 \begin{align}
G_{19}=-4g_2 \lambda^2 \mu^{3\epsilon}  \int \frac{d^d q}{(2\pi)^d} L_{\alpha\mu}(-p,q) D_{\mu\nu}(q) L_{\nu\beta}(-q,p) \frac{1}{(p-q)^4} \overset{\epsilon\rightarrow 0}{=}  \frac{2g_2 \lambda^2 }{3(4\pi)^3 \epsilon} L_{\alpha\beta} \label{m17} \ ,
\end{align}
\begin{align}
G_{20}&=8 \lambda^3 \mu^{3\epsilon}  \int \frac{d^d q}{(2\pi)^d} L_{\alpha\mu}(-p,q) D_{\mu\nu}(q) L_{\nu\rho}(-q,q)D_{\rho\sigma}(q) L_{\sigma\beta}(-q,p) \frac{1}{(p-q)^2} \overset{\epsilon\rightarrow 0}{=} \frac{4\lambda^3}{3(4\pi)^3\epsilon} L_{\alpha\beta} \label{m18} \ ,
\end{align}
where we introduced a shorthand notation $L_{\alpha\beta}\equiv L_{\alpha\beta}(p,-p)$. Notice that the graphs $G_{18}, G_{19}, G_{20}$ are one-loop corrections to the tree-level vertex $(\sigma -A -A)$ Tab. \ref{mT1} and hence they must be proportional to the same rank-2 tensor ($L_{\alpha\beta}$) as the tree-level vertex, which is confirmed by (\ref{m16}, \ref{m17}, \ref{m18}). The graph $G_{21}$ is zero in dimensional regularization in the massless limit. Demanding the counter-vertex $CV^{(\sigma A A)}$ (\ref{m24}) to cancel the divergencies coming from $G_{18}, G_{19}, G_{20}$ and using the values of $Z_\sigma$ and $Z_A$ from (\ref{m20}, \ref{m21}) we find the $( Z_{4u}, \ u=1,2,3,4)$
\begin{align}
&Z_{41}=\frac{N_f e^2}{6 (4\pi)^3 \epsilon}  \ ,\\
&Z_{42}=-\frac{\lambda^2}{3(4\pi)^3\epsilon} \ ,\\
&Z_{43}=0 \ , \\
&Z_{44}=1+\frac{N_f g_1^2}{12 (4\pi)^3 \epsilon} +\frac{g_2^2}{24 (4\pi)^3 \epsilon} -\frac{N_f e^2}{30 (4\pi)^3\epsilon}+\frac{11 \lambda^2}{6(4\pi)^3\epsilon} \ .
\end{align}
In principle we could have renormalized the vertex $A-\Phi-\Phi^*$ as well. However as we already remarked in the section \ref{HDG}, the gauge coupling is renormalized multiplicatively (\ref{m6}) and due to gauge invariance 
\begin{align}
Z_e=1/Z_A=1-\frac{N_f e^2}{60(4\pi)^3 \epsilon} \ .
\end{align}

 The final step is the construction of the beta functions with the help of the following equations 
 \begin{align} \label{eqforbeta}
 \epsilon g_u +\beta_{g_u}+ Z^{-1}_{u w} \frac{d Z_{w v}}{dg_h} g_v  \cdot \beta_{g_h}=0 \ , \ u=1,2,3,4
 \end{align}
 where summation over indices $w,v,h$ is assumed. The equations (\ref{eqforbeta}) follow from the Callan-Symanzik equations. Plugging values of the mixing matrix $Z_{uv}$ into (\ref{eqforbeta}) we find the beta functions\footnote{Notice that the one-loop beta function of the gauge coupling (\ref{m30}) is independent from the non-gauge couplings $(g_1,g_2,\lambda)$.  That the one-loop $\beta_e$ is independent from the $(g_1,g_2)$ was expected, since using those vertices one cannot draw a one-loop photon self-energy graph. On the other hand the photon self-energy graph $G_9$ vanishes, which would otherwise give a contribution $\sim e \lambda^2$ to the $\beta_e$. Similarly, in fermionic d=6 QCD \cite{Gracey:2015xmw}, the one-loop gauge coupling was proved to be independent from the non-gauge coupling. In that theory the non-gauge coupling stands for the interaction $\sim f_{ABC}G^A_{\alpha\beta} G^B_{\beta \gamma}G^C_{\gamma\alpha}$ (where the $G^A_{\alpha\beta},  \ A=1,...,N_c$ is the field strength of the $SU(N_c)$ gauge field), which unlike to the QED case doesn't vanish. See also \cite{Casarin:2019aqw}. We thank to John Gracey for drawing our attention to these points. }
 \begin{align} \label{m28}
 &\beta_{g_1}=-\epsilon g_1 + \frac{(N_f-4) g_1^3}{6 (4\pi)^3}- \frac{g_1^2 g_2}{ (4\pi)^3}+ \frac{g_1 g_2^2}{12 (4\pi)^3}- \frac{10 g_1 e^2}{3 (4\pi)^3}+ \frac{5 g_1 \lambda^2}{(4\pi)^3}+\frac{10 \lambda e^2}{(4\pi)^3}  \ ,\\ \label{m29}
 &\beta_{g_2}=-\epsilon g_2 -\frac{2N_f g_1^3}{(4\pi)^3}+\frac{N_f g_1^2 g_2}{2(4\pi)^3}-\frac{3g_2^3}{4(4\pi)^3}+\frac{15 g_2 \lambda^2}{(4\pi)^3}
 -\frac{40 \lambda^3}{(4\pi)^3}  \ ,\\ \label{m30}
 &\beta_e=-\epsilon e -\frac{N_f e^3}{30 (4\pi)^3} \ , \\ \label{m31}
 &\beta_\lambda=-\epsilon \lambda +\frac{N_f g_1 e^2}{3(4\pi)^3}-\frac{2g_2  \lambda^2}{3(4\pi)^3}+\frac{N_f g_1^2 \lambda}{6(4\pi)^3}
 +\frac{\lambda g_2^2}{12(4\pi)^3}-\frac{N_f \lambda e^2}{15(4\pi)^3}+\frac{11 \lambda^3}{3(4\pi)^3} \ .
 \end{align}
 
 \paragraph{Large $N_f$ limit of the beta functions:}

 We solve the beta functions (\ref{m28}, \ref{m29}, \ref{m30}, \ref{m31}) in the large $N_f$ limit. Besides the trivial fixed point where all the couplings vanish, we find three IR interacting fixed points.  One of the fixed points has a vanishing gauge coupling. It is the fixed point of the $O(2N_f)$-Yukawa theory \cite{Fei:2014yja}. The other two fixed points have a non-vanishing gauge coupling. We denote them as FP$_1$ and FP$_2$. The values of couplings at those fixed points are
  \begin{align} \nonumber
\text{FP}_1:   \\  \label{m32}
 g_1=&\sqrt{ \frac{6(4\pi)^3 \epsilon}{N_f}} \Big( 1+\frac{336}{N_f}+ O\Big(\frac{1}{N_f^2}\Big) \Big) \ , \\ \label{m33}
 g_2=&6  \sqrt{ \frac{6(4\pi)^3 \epsilon}{N_f}} \Big( 1+O\Big(\frac{1}{N_f}\Big) \Big) \ , \\ \label{m34}
 e^2=&-\frac{ 30 (4\pi)^3 \epsilon}{N_f}  \ ,\\ \label{m35}
 \lambda=&5 \sqrt{ \frac{6(4\pi)^3 \epsilon}{N_f}} \Big( 1+ O\Big(\frac{1}{N_f}\Big) \Big)  \ .\\
\nonumber  \text{FP}_2:   \\ \label{m35}
  g_1=&g_2=\lambda=0 \ , \\  \label{m36}
  e^2=&-\frac{ 30 (4\pi)^3 \epsilon}{N_f}  \ .
 \end{align}
Since at the fixed point FP$_2$ the couplings $(g_1, g_2, \lambda)$ vanish, the $\sigma$ field does not interact with any other field (including itself) and propagates freely. At the FP$_2$ the scalar flavors $\Phi_i$ are minimally coupled to the gauge field, with non-zero gauge coupling (\ref{m36}). The irrelevant $\Phi^4$ operator cannot be generated along the flow (at least if we are close to d=6). Therefore we can foresee that the FP$_2$ describes the critical scalar QED. Instead at the fixed point FP$_1$ neither of the couplings vanish (\ref{m32}, \ref{m33}, \ref{m34}, \ref{m35}) and it describes the critical $\bC\bP^{(N_f-1)}$.

 In order to test these statements, below we evaluate the scaling dimensions of the fields $(\Phi, \sigma, A)$ at the fixed points FP$_{1,2}$. Plugging in (\ref{m25}, \ref{m26}, \ref{m27}) the FP$_1$ values of the couplings we obtain  
 \begin{align} \label{m37}
& \Delta[\Phi_i]= 2-\epsilon +\frac{51\epsilon}{N_f}+O(\epsilon^2)  \ , \\  \label{m38}
 &\Delta[\sigma]=2+\frac{1440\epsilon}{N_f}+O(\epsilon^2) \ , \\  \label{m39}
 &\Delta[A_\mu]=1 \ .
 \end{align}
 The scaling dimension of the gauge field at the interacting fixed point is equal to 1 (actually this holds true at all orders in the perturbative expansion). We see a perfect match with the scaling dimensions of the fields $(\Phi, \sigma, A)$ calculated at the critical point $\bC\bP^{(N_f-1)}$ (\ref{vas9}, \ref{vas10}, \ref{vas11}) with the help of a large $N_f$ expansion.  
  
 Similarly, plugging in (\ref{m25}, \ref{m26}, \ref{m27}) the FP$_2$ values of the couplings we obtain  (as we have already mentioned the $\sigma$ field is free, and its scaling dimension is that of a free scalar field in $d=6-2\epsilon$)
 \begin{align} \label{m40}
 & \Delta[\Phi_i]= 2-\epsilon +\frac{50\epsilon}{N_f}+O(\epsilon^2) \ , \\ \label{m41}
 &\Delta[A_\mu]=1 \ .
 \end{align}
 Again, we find an agreement with the scaling dimensions of the fields $(\Phi, A)$ calculated at the critical pure scalar QED (\ref{vas14}, \ref{vas15}).

 \section{Renormalization of the mass parameters and the anomalous dimensions of the quadratic operators }

Until now we considered the theory (\ref{hard}) in the massless limit. When one turns on the masses, additional divergencies appear in the 2-point 1PI Green functions which must be cancelled with the appropriate mass counter-terms. The strategy for calculating these counter-terms is to first differentiate the 2-point Green functions with respect to the mass and then to put the mass equal to zero. In this way, quadratically divergent Green functions $\Gamma^{(\Phi\Phi^*)}$ and $\Gamma^{(\sigma\sigma)}$ become logarithmically divergent and the quarticly divergent Green function $\Gamma^{(AA)}$ becomes quadratically divergent.  The graphs which appear in the differentiated Green functions are collected in Tab. \ref{mT4}, \ref{mT5}, \ref{mT6}. We use the slash to mark the propagators which have been differentiated in a given graph. In the first lines of Tab. \ref{mT4}, \ref{mT5}, \ref{mT6} we provide the differentiated mass counter-terms, which are necessary for curing the divergencies. 

Using (\ref{ma1}) for the graph $G_{22}$ we obtain
\begin{align} \label{m42}
G_{22} = -g_1^2 \mu^{2\epsilon} \int \frac{d^d q}{(2\pi)^d} \frac{1}{q^4 (p-q)^2}\overset{\epsilon\rightarrow 0}{=}  -\frac{g_1^2}{2(4\pi)^3 \epsilon} \ .
\end{align}
The minus sign in front of the integral (\ref{m42}) comes from the differentiation of the scalar propagator $ \frac{\partial}{\partial\tau_1} \Big( \frac{1}{q^2+\tau_1} \Big) \Big|_{\tau_1=0}=-\frac{1}{q^4}$. The graph $G_{23}$ gives
\begin{align}\label{m43}
G_{23}=-e^2 \mu^{2\epsilon} \int \frac{d^d q}{(2\pi)^d} \frac{(2p-q)_\alpha (2p-q)_\beta}{(p-q)^4} D_{\alpha\beta}(q) =\text{finite} \ .
\end{align}
The absence of a pole in $G_{23}$  can be proved using the transversality condition $D_{\alpha\beta} q_{\alpha}=0$ in (\ref{m43}). Using (\ref{m42})  and the value of $Z_\phi$ from (\ref{m19}), we find  
\begin{align} 
Z_{11}^\tau=1-\frac{g_1^2}{3(4\pi)^3\epsilon}-\frac{5e^2}{3(4\pi)^3 \epsilon} \ .
\end{align}
The pole of the graph $G_{24}$ is the same as that of the $G_{22}$. Using it we find 
\begin{align}
Z_{12}^\tau=-\frac{g_1^2}{2(4\pi)^3\epsilon} \ .
\end{align}
The graph $G_{25}$ is finite (using the transversality condition). The tadpole $G_{26}$ gives
\begin{align} \label{m44}
G_{26}=e^2 \mu^{2\epsilon} \int \frac{d^d q}{(2\pi)^d}  \delta_{\alpha\beta} \frac{\delta_{\alpha\beta}-\frac{q_\alpha q_\beta}{q^2}}{(q^2+\tau_3)^3}\overset{\epsilon\rightarrow 0}{=}  \frac{5e^2}{2(4\pi)^3\epsilon} \ .
\end{align}
In the graph $G_{26}$, the gauge propagator is differentiated with respect to the gauge mass $\tau_3$.  In order to avoid the IR divergencies, in the integral (\ref{m44}) we kept a non-zero mass (which obviously doesn't effect the UV pole of the $G_{26}$). Using (\ref{m44}) we find
\begin{align}
Z_{13}^\tau=\frac{5e^2}{2(4\pi)^3 \epsilon} \ .
\end{align}
Notice that the loops in some of the graphs in Tab. \ref{mT2} are made by the propagators of the same field. Therefore differentiation will give two equivalent graphs with one propagator differentiated and the other one not. Since they are equivalent we simply multiply those graphs by two in Tab. \ref{mT4}, \ref{mT5}. The poles of the graphs $G_{27}$ and $G_{28}$ are extracted doing a calculation similar to the one in (\ref{m42}).  Using their values, which are recorded in Tab. \ref{mT4} and the value of $Z_\sigma$ we obtain 
\begin{align}
&Z_{21}^\tau=-\frac{N_f g_1^2}{(4\pi)^3 \epsilon} \ , \\
&Z_{22}^\tau=1+\frac{N_f g_1^2}{6 (4\pi)^3\epsilon}-\frac{5 g_2^2}{12(4\pi)^3 \epsilon}+\frac{5 \lambda^2}{(4\pi)^3 \epsilon} \ .
\end{align}
The graph $G_{29}$ gives 
\begin{align} \label{m45}
G_{29}=-4\lambda^2 \mu^{2\epsilon} \int \frac{d^d q}{(2\pi)^d} L_{\alpha\mu}(q,p-q) \frac{D_{\alpha\beta}(q)}{q^2} L_{\beta\nu}(q,p-q) D_{\mu\nu}(p-q)\overset{\epsilon\rightarrow 0}{=}  -\frac{10\lambda^2}{(4\pi)^3\epsilon} \ .
\end{align}
The minus sign in front of the integral (\ref{m45}) comes from the differentiation of the photon propagator (\ref{sixdimphoton}): $ \frac{\partial D_{\alpha\beta}(q)}{\partial\tau_3}  \Big|_{\tau_3=0}=-\frac{D_{\alpha\beta}(q)}{q^2}$. To extract the divergent part of the integral (\ref{m45}), it is sufficient to replace in it $L_{\alpha\mu}(q,p-q) \rightarrow -\delta_{\alpha\mu} q^2$ and $L_{\beta\nu}(q,p-q) \rightarrow -\delta_{\beta\nu} q^2$. This is because other terms inside these vertices either give finite contributions or vanish after multiplying them with photon propagators in (\ref{m45}). Using (\ref{m45}) we find 
\begin{align}
Z_{23}^\tau=-\frac{10\lambda^2}{(4\pi)^3 \epsilon} \ .
\end{align}
The graph $G_{30}$ gives 
\begin{align} \label{m46}
G_{30}= -2N_f e^2 \mu^{2\epsilon}\int \frac{d^d q}{(2\pi)^d}  \frac{(p+2q)_\alpha (p+2q)_\beta}{q^4 (q+p)^2} \overset{\epsilon\rightarrow 0}{=} -\frac{N_f e^2}{3(4\pi)^3 \epsilon} L_{\alpha\beta} \ .
\end{align}
The integral (\ref{m46}) is calculated introducing Feynman parametrization and using formulas (\ref{ma1}, \ref{ma4}). The tadpole $G_{31}$ is vanishing in the dimensional regularization. Using the (\ref{m46}) we obtain 
\begin{align}
Z_{31}^\tau=\frac{N_f e^2}{3(4\pi)^3 \epsilon} \ .
\end{align}
 The graphs $G_{32}$ and $G_{33}$ are different but it turns out that their poles are equal
\begin{align} \label{m47}
&G_{32}=-4 \lambda^2 \mu^{2\epsilon} \int \frac{d^d q}{(2\pi)^d}  \frac{1}{(p-q)^4} L_{\alpha\mu}(-p,q) D_{\mu\nu}(q) L_{\nu\beta}(-q,p)
 \overset{\epsilon\rightarrow 0}{=} \frac{2\lambda^2}{3(4\pi)^3 \epsilon} L_{\alpha\beta} \ , \\ \label{m48}
&G_{33}=-4\lambda^2 \mu^{2\epsilon}  \int \frac{d^d q}{(2\pi)^d} \frac{1}{(p-q)^2}  L_{\alpha\mu}(-p,q) \frac{D_{\mu\nu}(q)}{q^2 } L_{\nu\beta}(-q,p) \overset{\epsilon\rightarrow 0}{=}  \frac{2\lambda^2}{3(4\pi)^3 \epsilon} L_{\alpha\beta} \ .
\end{align}
Using (\ref{m47}, \ref{m48}) and the value of $Z_{A}$ from (\ref{m21}) we find
\begin{align}
&Z_{32}^\tau=-\frac{2\lambda^2}{3(4\pi)^3 \epsilon} \ , \\
&Z_{33}^\tau=1-\frac{N_f e^2}{30(4\pi)^3 \epsilon}-\frac{2\lambda^2}{3(4\pi)^3\epsilon} \ .
\end{align}
Having constructed the renormalization matrix $Z_{ab}^\tau$, which is responsible for the mixing between the masses (\ref{m4}), we now proceed to find the mixing matrix of the mass parameters. Those are defined as follows 
\begin{align} \label{m49}
\gamma_{ab}^\tau=\frac{d\ln Z_{ab}^\tau}{d\ln\mu}=(Z^{\tau})^{-1}_{ac} \frac{d Z_{cb}^\tau}{d g_v} \beta_{g_v}     \ ; \ \ \ \ a,b=1,2,3 \ \ ,
\end{align}
where summation over the indices $c=1,2,3$ and $v=1,2,3,4$ is assumed. Plugging in (\ref{m49}) the values of $Z^\tau$ matrix, we find 
\begin{align} \label{m50}
\gamma^\tau_{ab}= \frac{1}{(4\pi)^3} \begin{bmatrix}
    \frac{2g_1^2}{3}+\frac{10e^2}{3}       & g_1^2 & -5e^2 \\
   2N_f g_1^2       & -\frac{N_f g_1^2}{3}+\frac{5g_2^2}{6}-10\lambda^2 & 20\lambda^2 \\
   -\frac{2N_fe^2}{3} & \frac{4\lambda^2}{3} & \frac{N_f e^2}{15}+\frac{4\lambda^2}{3}
\end{bmatrix} \ ,
\end{align}
where we factored out the common factor $1/(4\pi)^3$. 
 \paragraph{The scaling dimensions of the mass operators at the fixed points:} We remind that the mixing matrix of the mass operators $(\Phi^2, \sigma^2, F_{\alpha\beta}^2)$ is minus the  (\ref{m50}). This is because the sum of the scaling dimensions of the mass and of the mass operator should be equal to $d=6-2\epsilon$ and we know that the classical dimensions of the mass and of the mass operator are respectively 2 and $4-2\epsilon$. 
 
 First, let us construct the mixing matrix of the mass parameters at the fixed point FP$_1$. Plugging in (\ref{m50}) the FP$_1$ values of the couplings (\ref{m32}, \ref{m33}, \ref{m34}, \ref{m35}) and keeping the entries of the matrix to the order $1/N_f$ we find
  \begin{align} \label{m51}
\gamma^\tau_{ab}\bigg|_{\text{FP$_1$}}=\epsilon \begin{bmatrix}
    \frac{-96}{N_f}       & \frac{6}{N_f} & \frac{150}{N_f} \\
   12\big(1+\frac{672}{N_f}\big)       & -2\big(1+\frac{1332}{N_f}\big) & \frac{3000}{N_f} \\
   20 & \frac{200}{N_f} & -2\big(1-\frac{100}{N_f} \big)
\end{bmatrix} \ .
\end{align}
The eigenvalues of the matrix (\ref{m51}), taken with an opposite sign are the anomalous scaling dimensions of the mass operators\footnote{To be more precise we should refer to the mass eigenstates rather than to the mass operators, since after diagonalization of the matrix (\ref{m51}) the operators ($\Phi^2,\sigma^2, F_{\alpha\beta}^2$) mix with each other.}. The full scaling dimensions are as follows 
\begin{align} \label{m52}
&\Delta_1^{(\text{FP$_1$})}= 4-2\epsilon- \Big(-2\epsilon-\frac{2000+280\sqrt{10}}{N_f}\epsilon\Big)+O(\epsilon^2)=4+\frac{40(50+7\sqrt{10})}{N_f}\epsilon+O(\epsilon^2) \ , \\ \label{m53}
&\Delta_2^{(\text{FP$_1$})}= 4-2\epsilon- \Big(-2\epsilon-\frac{2000-280\sqrt{10}}{N_f}\epsilon \Big)+O(\epsilon^2)=4+\frac{40(50-7\sqrt{10})}{N_f}\epsilon+O(\epsilon^2) \ , \\ \label{m54}
&\Delta_3^{(\text{FP$_1$})}=4-2\epsilon-\frac{1440\epsilon}{N_f}+O(\epsilon^2) \ .
\end{align}
 Again we find a perfect agreement with the scaling dimensions of these operators at the critical point $\bC\bP^{(N_f-1)}$ (\ref{vas19}, \ref{vas12}, \ref{vas13}).
 
  Finally, let us plug in (\ref{m50}) the FP$_2$ values of the couplings to determine the anomalous mixing matrix of the mass parameters at that fixed point 
  \begin{align} \label{m55}
 \gamma^\tau_{ab}\bigg|_{\text{FP}_2}= \epsilon \begin{bmatrix}
    -\frac{100}{N_f}       & \frac{150}{N_f} \\
   20 & -2
   \end{bmatrix} \ .
 \end{align}
 The eigenvalues of the matrix (\ref{m55}), taken with an opposite sign are the anomalous scaling dimensions of the mass operators ($\Phi^2, F_{\alpha\beta}^2$). The full scaling dimensions are as follows 
 \begin{align} \label{m56}
 &\Delta_1^{(\text{FP$_2$})}=4-2\epsilon-\frac{1400}{N_f} \epsilon+O(\epsilon^2)  \ ,\\ \label{m57}
 &\Delta_2^{(\text{FP$_2$})}=4+\frac{1500}{N_f} \epsilon+O(\epsilon^2) \ .
 \end{align}
 One of the eigenvalues ($\Delta_1^{(\text{FP$_2$})}$) matches with the scaling dimension of the $\Phi^2$ operator calculated at the critical pure scalar QED (\ref{vas18}). We do not have a formula for the scaling dimension (order  $O(1/N_f)$) of the $F_{\alpha\beta}^2$ operator at the critical pure scalar QED, and so we cannot provide a check for (\ref{m57}). We remind that at the fixed point FP$_2$ the $\sigma$ field doesn't interact, therefore the scaling dimension of the operator $\sigma^2$ is simply twice a scaling dimension of a free scalar field.

 \begin{table}[]
 \centering 
 \begin{tikzpicture} [scale=0.6]  
 
 \draw[] (-1,5) to (25.5, 5);
 \draw[] (-1,3.5) to (25.5,3.5);
 \draw[] (-1,5) to (-1,-4);
 \draw[] (-1,-4) to (25.5,-4);
 \draw[] (25.5,5) to (25.5,-4);
 \draw[] (8.5,5) to (8.5,-4);
  \draw[] (17,5) to (17,-4);
  
  \node at (4,4.2) {$\partial_{\tau_1}\Gamma^{(\Phi\Phi^*)}|_{\tau_1=\tau_2=\tau_3=0}$};
  \node at (13,4.2) {$\partial_{\tau_2}\Gamma^{(\Phi\Phi^*)}|_{\tau_1=\tau_2=\tau_3=0}$};
  \node at (21,4.2) {$\partial_{\tau_3}\Gamma^{(\Phi\Phi^*)}|_{\tau_1=\tau_2=\tau_3=0}$};

\node at (0,2) {$\frac{\partial}{\partial \tau_1} \Big($};
\node at (2.6,2) {$ \Big)$};

 \draw[thick] (0.5,2) to (2.5,2);
 \node at (5.7,2) {$=  -(Z_\Phi^2 Z_{11}^\tau-1)$};
\draw[fill] (1.5,2) circle (4pt);   
 \node at (-0.2,0.2) {$\scriptstyle{G_{22}}$};
\draw[thick] (0,-0.5) to (3,-0.5);
\draw[dashed, thick] ([shift=(180:1cm)]1.5,-0.5) arc (180:0:1 cm);
\node at (1.5,-0.5) {$/$};
\node at (5.3,-0.3) {$=  \ -\frac{g_1^2}{2(4\pi)^3 \epsilon}$};
 \node at (-0.2,0.2-2.5) {$\scriptstyle{G_{23}}$};
\draw[thick] (0,-3) to (3,-3);
\node at (1.5,-3) {$/$};
\draw [wavy semicircle, thick] (0.5,-3 ) to  (2.5,-3);
\node at (4.6,-2.8) {$=  \ 0$};

\begin{scope}[shift={(9.7,0)}]
\node at (0,2) {$\frac{\partial}{\partial \tau_2} \Big($};
\node at (2.6,2) {$ \Big)$};
\draw[thick] (0.5,2) to (2.5,2);
 \node at (4.9,2) {$= -Z_\Phi^2 Z_{12}^\tau$};
\draw[fill] (1.5,2) circle (4pt);  
 \node at (-0.2,0.2) {$\scriptstyle{G_{24}}$}; 
\draw[thick] (0,-0.5) to (3,-0.5);
\draw[dashed, thick] ([shift=(180:1cm)]1.5,-0.5) arc (180:0:1 cm);
\node at (1.5,0.5) {$/$};
\node at (5.3,-0.3) {$=  \ -\frac{g_1^2}{2(4\pi)^3 \epsilon}$};
\end{scope}

\begin{scope}[shift={(18.4,0)}]
\node at (0,2) {$\frac{\partial}{\partial \tau_3} \Big($};
\node at (2.6,2) {$ \Big)$};
\draw[thick] (0.5,2) to (2.5,2);
 \node at (5,2) {$= -Z_\Phi^2Z_{13}^\tau$};
\draw[fill] (1.5,2) circle (4pt);   
 \node at (-0.2,0.2) {$\scriptstyle{G_{25}}$};
\draw[thick] (0,-0.5) to (3,-0.5);
\draw [wavy semicircle, thick] (0.5,-0.5) to  (2.5,-0.5);
\node at (4.5,-0.3) {$= \ 0$};
 \node at (-0.2,0.2-2.5) {$\scriptstyle{G_{26}}$};
\node at (1.5,0.5) {$/ $};
\node at (5.3,-2.8) {$= \ \frac{5e^2 }{2(4\pi)^3 \epsilon}$};
 \draw[thick] (0,-3) to (3,-3);
 \draw[thick,decorate,decoration={complete siness}] (0.9,-2.26) arc (180-5:-180-5:0.6);
 \node at (1.5,-1.7) {$/ $};

\end{scope}

 \end{tikzpicture}
\caption{ 2-point Green function $\Gamma^{(\Phi\Phi^*)}$ differentiated w.r.t masses}  \label{mT4}
\end{table}

 \begin{table}[]
 \centering 
 \begin{tikzpicture} [scale=0.6]  
 
 \draw[] (-1,5) to (25.5, 5);
 \draw[] (-1,3.5) to (25.5,3.5);
 \draw[] (-1,5) to (-1,-2);
 \draw[] (-1,-2) to (25.5,-2);
 \draw[] (25.5,5) to (25.5,-2);
 \draw[] (7.5,5) to (7.5,-2);
  \draw[] (17,5) to (17,-2);
  
  \node at (3.7,4.2) {$\partial_{\tau_1}\Gamma^{(\sigma \sigma)}|_{\tau_1=\tau_2=\tau_3=0}$};
  \node at (13,4.2) {$\partial_{\tau_2}\Gamma^{(\sigma \sigma)}|_{\tau_1=\tau_2=\tau_3=0}$};
  \node at (21,4.2) {$\partial_{\tau_3}\Gamma^{(\sigma\sigma)}|_{\tau_1=\tau_2=\tau_3=0}$};

\node at (0,2) {$\frac{\partial}{\partial \tau_1} \Big($};
\node at (2.6,2) {$ \Big)$};

 \draw[thick,dashed] (0.5,2) to (2.5,2);
 \node at (5.3,2) {$=  -Z_\sigma^2 Z_{21}^\tau$};
\draw[fill] (1.5,2) circle (4pt);   
 \node at (1.5,-0.6) {$\scriptstyle{N_f}$};
 \node at (0.1,0.2-1.5) {$\scriptstyle{G_{27}}$};
\draw[dashed,thick] (0,-0.5) to (0.8,-0.5);
\draw[thick] ([shift=(180:1cm)]1.8,-0.5) arc (180:-180:0.7 cm);
\draw[dashed,thick] (2.2,-0.5) to (3.03,-0.5);
\node at (1.5,0.2) {$/$};
\node at (-0.5,-0.5) {\scalebox{0.8}{$ 2 \times$}};
\node at (5,-0.5) {$=-\frac{N_f g_1^2}{(4\pi)^3 \epsilon}$};

\begin{scope}[shift={(8.7,0)}]
\node at (0,2) {$\frac{\partial}{\partial \tau_2} \Big($};
\node at (2.6,2) {$ \Big)$};
\draw[thick, dashed] (0.5,2) to (2.5,2);
 \node at (5.5,2) {$= -(Z_\sigma^2 Z_{22}^\tau-1)$};
\draw[fill] (1.5,2) circle (4pt);  
 \node at (0.1,0.2-1.5) {$\scriptstyle{G_{28}}$};
\draw[dashed,thick] (0,-0.5) to (0.8,-0.5);
\draw[thick,dashed] ([shift=(180:1cm)]1.8,-0.5) arc (180:-180:0.7 cm);
\draw[dashed,thick] (2.2,-0.5) to (3.03,-0.5);
\node at (1.5,0.2) {$/$};
\node at (-0.5,-0.5) {\scalebox{0.8}{$ 2 \times$}};
\node at (5,-0.5) {$=-\frac{ g_2^2}{2(4\pi)^3 \epsilon}$};
\end{scope}

\begin{scope}[shift={(18.4,0)}]
\node at (0,2) {$\frac{\partial}{\partial \tau_3} \Big($};
\node at (2.6,2) {$ \Big)$};
 \node at (0.1,0.2-1.5) {$\scriptstyle{G_{29}}$};
\draw[thick,dashed ] (0.5,2) to (2.5,2);
 \node at (5,2) {$= -Z_\sigma^2Z_{23}^\tau$};
\draw[fill] (1.5,2) circle (4pt);   

\begin{scope}[shift={(-0.5,0.4)}]
\node at (1.5,-0.3) {$/$};
\node at (-0.4,-1) {\scalebox{0.8}{$ 2 \times$}};
\draw[dashed,thick] (0,-1) to (0.9,-1);
 \draw[thick,decorate,decoration={complete siness}] (0.9,-3.65+2.8) arc (180-5:-180-5:0.6);
 \draw[dashed,thick] (2.2,-1) to (3.03,-1);
\end{scope}
\node at (5,-0.5) {$=-\frac{10 \lambda^2}{(4\pi)^3 \epsilon}$};
\end{scope}

 \end{tikzpicture}
\caption{ 2-point Green function $\Gamma^{(\sigma \sigma)}$ differentiated w.r.t masses}  \label{mT5}
\end{table}

 \begin{table}[]
 \centering 
 \begin{tikzpicture} [scale=0.6]  
 
 \draw[] (-1,5) to (26.5, 5);
 \draw[] (-1,3.5) to (26.5,3.5);
 \draw[] (-1,5) to (-1,-4);
 \draw[] (-1,-4) to (26.5,-4);
 \draw[] (26.5,5) to (26.5,-4);
 \draw[] (8.5,5) to (8.5,-4);
  \draw[] (17,5) to (17,-4);
  
  \node at (4,4.2) {$\partial_{\tau_1}\Gamma^{(A A)}|_{\tau_1=\tau_2=\tau_3=0}$};
  \node at (13,4.2) {$\partial_{\tau_2}\Gamma^{(A A)}|_{\tau_1=\tau_2=\tau_3=0}$};
  \node at (21,4.2) {$\partial_{\tau_3}\Gamma^{(A A)}|_{\tau_1=\tau_2=\tau_3=0}$};

\node at (0,2) {$\frac{\partial}{\partial \tau_1} \Big($};
\node at (2.6,2) {$ \Big)$};

 \draw[photon2] (0.5,2) to (2.5,2);
 \node at (5.7,2) {$=  Z_A^2 Z_{31}^\tau L_{\alpha\beta}$};
\draw[fill] (1.5,2) circle (4pt);   

\begin{scope}[shift={(-0.2,1.8)}]
\node at (1.7,-1.3) {$/$};
\node at (-0.3,-2) {\scalebox{0.8}{$ 2 \times$}};
 \draw[photon3] (0,-2) to (1,-2);
 
 \draw[ thick] ([shift=(180:1cm)]2,-2) arc (180:-180:0.7 cm);
 \draw[photon3] (2.4,-2) to (3.4,-2);
 \node at (1.7,-2.1) {$\scriptstyle{N_f}$};
  \node at (0.5,0.2-3) {$\scriptstyle{G_{30}}$};
   \node at (6.2,-2) {$= \ -\frac{N_f e^2}{3(4\pi)^3\epsilon}L_{\alpha\beta}$};
     \node at (0.5,0.1-5.5) {$\scriptstyle{G_{31}}$};
    \draw[photon2] (0,-4.8) to (1.7,-4.8);
    \draw[photon2] (1.7,-4.8) to (3.4,-4.8);
     \draw[thick] (1.75,-4.13) circle (0.6cm); 
     \node at (1.7,-3.6) {$/$};
     \node at (5,-4.6) {$=0$};
   \end{scope}
   
\begin{scope}[shift={(9.7,0)}]
\node at (0,2) {$\frac{\partial}{\partial \tau_2} \Big($};
\node at (2.6,2) {$ \Big)$};
\draw[photon2] (0.5,2) to (2.5,2);
 \node at (4.9,2) {$= Z_A^2 Z_{32}^\tau L_{\alpha\beta}$};
   \node at (0.2,-1.2) {$\scriptstyle{G_{32}}$};
\draw[fill] (1.5,2) circle (4pt);   
 \draw[photon2] (-0.5,-0.5) to (3.4-0.5,-0.5);
   \draw[dashed, thick] ([shift=(180:0.9cm)]1.8-0.5,-0.4) arc (180:0:0.9 cm);
   \node at (1.3,0.5) {$/$};
   \node at (5.1,-0.4) {$=\frac{2\lambda^2}{3(4\pi)^3 \epsilon} L_{\alpha\beta}$};
\end{scope}

\begin{scope}[shift={(18,0)}]
\node at (0,2) {$\frac{\partial}{\partial \tau_3} \Big($};
\node at (2.6,2) {$ \Big)$};
\draw[photon2] (0.5,2) to (2.5,2);
 \node at (5.7,2) {$= (Z_A^2Z_{33}^\tau-1) L_{\alpha\beta}$};
\draw[fill] (1.5,2) circle (4pt);  
  \node at (0.2,-1.2) {$\scriptstyle{G_{33}}$}; 
 \draw[photon2] (-0.5,-0.5) to (3.4-0.5,-0.5);
   \draw[dashed, thick] ([shift=(180:0.9cm)]1.8-0.5,-0.4) arc (180:0:0.9 cm);
 \node at (1.4,-0.4) {$/$};
   \node at (5.2,-0.4) {$=\frac{2\lambda^2}{3(4\pi)^3 \epsilon} L_{\alpha\beta}$};

\end{scope}

 \end{tikzpicture}
\caption{2-point Green function $\Gamma^{(A A)}$ differentiated w.r.t masses}  \label{mT6}
\end{table}

\section{Conclusion and outlook}
In this paper we studied the critical $\bC\bP^{(N_f-1)}$ NLSM and the critical pure scalar QED in dimension $4<d<6$.  We proved, that these critical points can be thought as IR fixed points of the HDG theory (\ref{hard}). We want to mention possible directions for future studies.

 In this paper, we have solved the beta functions (\ref{m28}, \ref{m29}, \ref{m30}, \ref{m31}) near $d=6$ only when the number of flavors is large. It would be interesting to further study these beta functions and the RG flow diagram near 6 dimensions, when the number of flavors $N_f$ is small.  We didn't study the higher loop corrections to the beta functions and to the anomalous dimensions. It will be interesting to study the effects of the 2-loop order corrections. It seems to us, that it will be easier to start with HDG action (\ref{hard}) by giving a mass to $\sigma$ and decoupling it from the spectrum. In this case one will need only to renormalize the gauge coupling. 
 
 In the paper \cite{Benvenuti:2018cwd}, using the marginality crossing equations  \cite{Gukov:2016tnp} and the large $N_f$ scaling dimensions for various quartic and quadratic operators,  it was argued that in d=3 the scalar QED (with $\Phi^4$ interaction) merges and annihilates with the tricritical scalar QED ($\Phi^4$ operator tuned to zero) at $N_f^* \sim 9-10$ (the $N_f^*$ is the critical number of flavors for which the QED's collide). Repeating the exercise for the $\bC\bP^{(N_f-1)}$ NLSM we obtain 
 \begin{align}
 &\Delta[-\frac{9+\sqrt{91}}{80}\sigma^2+F^{\mu\nu}F_{\mu\nu}]=4+\frac{512(43+3\sqrt{91})}{15 \pi^2 N_f} =5  \\
 & \ \ \ \ \ \ \ \ \ \ \ \ \  \ \ \ \ \ \ \ \ \ \  \ \ \Rightarrow N_f^* =\frac{512(43+3\sqrt{91})}{15\pi^2}\sim 247.68  \  \ \ . \label{verybig}
 \end{align}
It is surprising that the critical number of flavors in $d=5$ is so big as compared to the $d=3$ case. Using the bootstrap methods \cite{Nakayama:2014yia, Chester:2014gqa} one might be able to check the (\ref{verybig}).
 
 For the bootstrap studies it might be useful to construct the coefficient $C_T$ of the 2-point function of the energy-momentum tensor in the large $N_f$ limit. To our knowledge, the $C_T$  for the d-dimensional critical $\bC\bP^{(N_f-1)}$ is not known yet\footnote{See \cite{Osborn:2016bev}, where the $C_T$ was obtained for higher dimensional higher derivative free field theories.}. 

\acknowledgments{ I am grateful to Sergio Benvenuti for many useful discussions about this and related topics. 
I would like to thank  Anzhela Sargsyan for support and encouragement.  Also I would like to thank John Gracey for useful email correspondence.  }

\appendix

\section{Useful formulae} \label{usefulformulas}

\begin{align}& \text{Feynman parametrization: }
\nonumber  \\ \label{ma3}
  & \ \ \ \  A_1^{\alpha_1} ...  A_n^{\alpha_n} = \frac{\Gamma\big(\sum\limits_{i=1}^{n} \alpha_i\big)}{\prod\limits_{i=1}^n \Gamma\big(\lambda_i\big)} \int_0^1 dx_1 ...  \int_0^1 dx_n \frac{\delta\big( \sum\limits_{i=1}^n x_i-1 \big) \prod\limits_{i=1}^n x_i^{\lambda_i-1}} {\Big[\sum\limits_{i=1}^n A_i x_i\Big]^{\sum \lambda_i}} \ , \\~ \nonumber \\
 \label{ma2}
&\int \frac{d^d q}{(2\pi)^d} \frac{1}{(q-p)^{2\alpha} q^{2\beta}}=\frac{\Gamma(\frac{d}{2}-\alpha)  \Gamma(\frac{d}{2}-\beta) \Gamma(\alpha+\beta-\frac{d}{2}) }{(4\pi)^{d/2} \Gamma(\alpha) \Gamma(\beta) \Gamma(d-\alpha-\beta) } p^{d-2\alpha-2\beta} \ ,  \\~  \nonumber \\
 \label{ma1}
&\int \frac{d^d q}{(2\pi)^d} \frac{1}{(q^2+\tau)^\alpha q^{2\beta}} = \frac{\Gamma(\alpha+\beta-\frac{d}{2}) \Gamma(\frac{d}{2}-\beta)}{(4\pi)^{d/2} \Gamma(\alpha) \Gamma(\frac{d}{2})} \tau^{d/2-\alpha-\beta}  \ , \\~  \nonumber \\
 \label{ma4}
  & \int \frac{d^d q}{(2\pi)^d}  f(q^2) q_\mu q_\nu = \frac{\delta_{\mu\nu}}{d} \int \frac{d^d q}{(2\pi)^d} f(q^2) q^2  \ ,\\~ \nonumber  \\
  \label{ma5}
   & \int \frac{d^d q}{(2\pi)^d}  f(q^2) q_\mu q_\nu q_\rho q_\sigma= \frac{\delta_{\mu\nu} \delta_{\rho\sigma}+\delta_{\mu\rho} \delta_{\nu\sigma}+\delta_{\mu\sigma} \delta_{\nu\rho}}{d(d+2)} \int \frac{d^d q}{(2\pi)^d} f(q^2) q^4 \ .
\end{align}

\bibliographystyle{ytphys}

\end{document}